\documentclass[11pt,tightenlines,eqsecnum,floats,aps,amsmath,amssymb,nofootinbib,prd,shownopac,floatfix]{revtex4-1}

\usepackage{graphicx}
\usepackage{epstopdf}
\usepackage{latexsym}
\usepackage{amssymb}
\usepackage{amsmath}
\usepackage{color}
\usepackage{mathrsfs}
\usepackage{xparse}
\usepackage{float}
\usepackage{comment}
\usepackage{mathtools}
\usepackage{multirow}
\usepackage{hyperref}
\usepackage[center]{subfigure}
\usepackage{tensor}
\usepackage{physics}
\usepackage{amsthm}
\usepackage{ulem} %

\usepackage{xargs}                      %

\allowdisplaybreaks

\makeatletter
\renewcommand{\p@subsection}{}
\renewcommand{\p@subsubsection}{}
\makeatother

\newtheorem{innercustomgeneric}{\customgenericname}
\providecommand{\customgenericname}{}
\newcommand{\newcustomtheorem}[2]{%
  \newenvironment{#1}[1]
  {%
   \renewcommand\customgenericname{#2}%
   \renewcommand\theinnercustomgeneric{##1}%
   \innercustomgeneric
  }
  {\endinnercustomgeneric}
}

\newcustomtheorem{customthm}{Theorem}
\newcustomtheorem{customlemma}{Lemma}
\newcustomtheorem{customcorollary}{Corolllary}
\newcustomtheorem{customproof}{Proof}

\newcommand{\mubar}{\overline{\mu}}

\begin{document}

\title{Dynamically implementing the $\mubar$-scheme in cosmological and spherically symmetric models in an extended phase space model}
\author{Kristina Giesel}
\email{kristina.giesel@gravity.fau.de}
\author{Hongguang Liu}
\email{hongguang.liu@gravity.fau.de}
\affiliation{Institute for Quantum Gravity,  Department of Physics, Theoretical Physics III, FAU Erlangen-N\"urnberg, Staudtstr. 7, 91058 Erlangen, Germany}

\begin{abstract}
We consider an extended phase space formulation for cosmological and spherically symmetric models in which the choice of a given $\mubar$-scheme can be implemented dynamically. These models are constructed in the context of the relational formalism by using a canonical transformation on the extended phase space which provides a Kucha\v{r} decomposition of the extended phase space. The resulting model can be understood as a gauge-unfixed model of a given $\mubar$-scheme. We use this formalism to investigate the restrictions to the allowed $\mubar$-scheme from this perspective and discuss the differences in the cosmological and spherically symmetric case. This method can be useful, for example, to obtain a $\mubar$-scheme in a top-down derivation from full LQG to symmetry reduced effective models, where for some models only the $\mu_0$-scheme has been obtained so far.
\end{abstract}

\maketitle
\section{Introduction}
The investigation of symmetry reduced models for loop quantum gravity (LQG) in the context of cosmological or spherically symmetric spacetimes is a topic of growing interest in recent years, see for spherically symmetric models for instance \cite{Ashtekar:2005qt,Modesto:2005zm,Boehmer:2007ket,Chiou:2012pg,Gambini:2013hna,Brahma:2014gca,Corichi:2015xia,Dadhich:2015ora,Tibrewala:2013kba,BenAchour:2017ivq,Yonika:2017qgo,DAmbrosio:2020mut,Olmedo:2017lvt,Ashtekar:2018lag,Ashtekar:2018cay,Bojowald:2018xxu,BenAchour:2018khr,Bodendorfer:2019cyv,Alesci:2019pbs,Assanioussi:2019twp,Benitez:2020szx,Gan:2020dkb,Gambini:2020qhx,Husain:2022gwp,Li:2021snn,Gan:2022mle,Kelly:2020lec,Gambini:2020nsf,Han:2020uhb,Zhang:2021xoa,Han:2022rsx,Munch:2022teq} or \cite{Ashtekar:2023cod} for a recent review  and \cite{Agullo:2016tjh} for a review on loop quantum cosmology (LQC). In this context effective models for LQC or quantum black holes are a useful framework to obtain first insights on the underlying symmetry reduced quantum model and its characteristic properties. A common feature such effective models share is that holonomy operators involved in full LQG are replaced by so called polymerization functions that mimic the loop inspired quantization of these symmetry reduced models. Such a polymerization function is always accompanied by a so-called polymerization parameter, usually denoted by $\mu$, which is the relevant scale for e.g. holonomy corrections in these effective models. The first formulation of such models used a fixed polymerization parameter that is independent of the phase space variables, the so called $\mu_0$-scheme. The usage of a fixed parameter is inspired by the regularization of holonomies of connections used in full loop quantum gravity \cite{QSD}. One problem of the $\mu_0$-scheme in effective LQC dynamics is that the critical density at the bounce depends on the initial values and is possibly non-Planckian. In the context of the so called  improved dynamics a $\mubar$-scheme  for LQC was then proposed to overcome this problem \cite{Ashtekar:2006rx}, in which the polymerization parameter is related to the minimal area gap and thus depends on the triad operators.
Going beyond cosmological models the improved dynamics were then proposed for spherically symmetric models \cite{Chiou:2012pg,Tibrewala:2013kba,BenAchour:2017ivq,Ashtekar:2018lag,Gambini:2020nsf,Kelly:2020uwj,Han:2020uhb,Zhang:2021xoa,Han:2022rsx}. We note that there are different implementations of the improved dynamics in these models, unlike in LQC where the choice is unique if we require scale invariance. 

To establish a link between full LQG and the LQC effective dynamics, a model for a coherent state path integral for LQG was introduced in  \cite{Han:2019vpw}. However, obtaining the improved $\overline{\mu}$-scheme from such top-down derivations of full LQG is still an open problem \cite{Dapor:2019mil,Han:2019vpw}. If one generalises the models in \cite{Dapor:2019mil,Han:2019vpw} to the extended phase space introduced here we expect that these models yield corresponding effective models with a $\overline{\mu}$-scheme in cosmology or spherically symmetric models. Thus, in these top-down derivations, the difference is whether one works with or without the extended phase space to obtain a model in either a $\mu_0$- or $\overline{\mu}$-scheme. There are some attempts to generalise models in a different directions \cite{Alesci:2016rmn,Han:2019feb,Han:2021cwb} to obtain effective models with a $\overline{\mu}$-scheme which involve either a different regularization of the dynamics or an ensemble of microstates labeled by different graphs. The approach to the implementation of the $\overline{\mu}$-scheme in this work is more along the lines introduced in the effective cosmological models, namely as a canonical transformation on the phase space of gravitational degrees of freedom.

In this paper, we aim at constructing a method based on an extended phase space that allows the implementation of a general $\mubar$-scheme dynamically and would like to investigate the restrictions to the allowed $\mubar$-scheme. We will discuss the model in the framework of the relational formalism \cite{Rovelli:1990ph,Rovelli:1993bm,Dittrich:2004cb,Dittrich:2005kc,Thiemann:2004wk} where we can formulate the model in terms of Dirac observables with respect to the constraints related to the choice of a given $\mubar$-scheme. The construction of such Dirac observables allows to define a canonical transformation on the extended phase space which maps the original kinematical variables into a new set of variables where the constraints related to the $\mubar$-scheme become new elementary variables in phase space. By this we achieve a so called Kucha\v{r}-decomposition \cite{Kuchar:1972,Hajicek:1999ht} of the phase space in which the constraints as well as the Dirac observables with respect to these constraints become the new elementary phase space variables. Working in extended phase space allows the choice of a particular $\mubar$-scheme to be implemented as a canonical transformation. Although this is also possible directly in the reduced phase space of cosmological models, it is no longer possible in reduced phase space for general ${\mubar}$-schemes in spherically symmetric models.

With the detailed construction of the cosmological model in Sec. \ref{sec:cos} and spherically symmetric model in Sec. \ref{sec:BH}, we show that the model in terms of these Dirac observables can be understood as the gauge-unfixed version of a given $\mubar$ model. The model includes constraints that fix $\mubar$ to some chosen phase space functions, that is implementing a specific $\mubar$-scheme, and we will call such functions gauge fixing functions in the following.  By requiring that the elementary canonical variables encoding the constraints vanish, we obtain the Hamiltonian constraint in the $\mubar$-scheme characterized by the choice of the gauge fixing functions, which can be taken as the starting point to further study of the quantum theory and effective dynamics. In addition we also investigate the corresponding gauge  fixed model for both the cosmological and spherically symmetric case. The method we propose here can for instance be applied to the path integral formulation of full LQG \cite{Han:2019vpw}, allowing to obtain $\mubar$-like effective dynamics from the top-down derivations of full LQG.

As our analysis shows in the cosmological model, the gauge fixing function encoded in the model, and thus the choice for $\mubar$, has no restrictions to the form of the $\mubar$ function. Any gauge fixing function that one chooses, which depends on the scale factor  and extrinsic curvature can lead to a well-defined Hamiltonian in the partially reduced phase space equipped with the standard canonical commutation relations (CCR). The form of the $\mubar$-scheme is fixed only when we further require scale invariance of the polymerization function, which then reduces to the standard $\mubar$-scheme used in LQC. In the spherically symmetric model which is a $1+1$-dimensional field theory, this is no longer the case. Due to the requirement for a consistent density weight of the effective constraints, the  dynamical implementation of a chosen $\mubar$-scheme is only possible if the polymerization depends on a combination of the extrinsic curvature and the triad such that they have density weight zero. This gives a restriction to the possible gauge fixing functions for spherical symmetric models. We note that if one only works with the interior of a black hole which can be described as a homogeneous Kantowski-Sachs spacetime, the gauge fixing functions are again free from any restriction similar to the case of cosmological models.

The paper is structured as follows: in section \ref{sec:cos} we present the dynamical implementation of the $\mubar$-scheme for cosmological models. In subsection \ref{sec:DiracObsCanTrafo} we use the framework of the relational formalism and discuss in detail how a canonical transformation can be constructed. Using the transformed variables we show on the one hand that we obtain a gauge-unfixed model for each given choice of a $\mubar$-scheme, including also the special choice of $\mu_0$, and on the other hand 
how the reduced phase space can be obtained. In subsection \ref{sec:GFmodel} we present the corresponding gauge fixed model for the cosmological case. In section \ref{sec:BH} we generalize our strategy to spherically symmetric models where in general more than one polymerization parameter is present and we consider field theory models. The sections on the cosmological and the spherically symmetric models should be self-contained, so that the reader interested in only one of the two cases can go directly to the appropriate section. As shown in subsection \ref{sec:DiracObsCanTrafoBH} a similar canonical transformation on the extended phase space can be constructed using the relational formalism. To show that the transformation is indeed canonical requires a little bit more effort compared to the cosmological case but works analogously. Again the resulting model can be understood as a gauge-unfixed model of a given $\mubar$-scheme. In addition we present in subsection \ref{sec:GFmodelBH} the corresponding gauge-fixed model. We summarize and conclude in section \ref{sec:GFmodelBH} and provide further details for some calculations in the appendix.

\section{Implementing the $\mubar$-scheme in cosmology}\label{sec:cos}
We aim at constructing a model on an extended phase space that allows to implement the $\mubar$-scheme in cosmology dynamically. 
Note that we will not consider a further gauge fixing or the construction of Dirac observables with respect to the Hamiltonian constraint but see the model presented here rather as a starting point for a cosmological model with a given Hamiltonian constraints for which on the usual kinematical phase space a $\mubar$-scheme needs to be chosen. Then, afterwards one can still consider a further gauge fixing of the Hamiltonian constraint or construct the fully reduced phase space that also involves the Hamiltonian constraint because usually such a step is done within an already given $\mubar$-scheme.
\subsection{Extended phase space for cosmology: Canonical Transformation and Dirac Observables}
\label{sec:DiracObsCanTrafo}
We consider the usual kinematical phase space of cosmology with elementary phase space variables denoted by $\theta,p,N,p_N,\phi^A,\pi_A$, where $N$ is the lapse function, $\theta,p$ the gravitational variables and $\phi^A,\pi_A$  some generic matter degrees of freedom where $A\in 1,\cdots, L_{\#}$ is a generic label to allow more than one matter component in general. We extend this phase space by two additional canonical pairs $(\lambda,p_\lambda)$ and $(\mu,p_\mu)$.  The non-vanishing CCR in the $(8+2L_{\#})$ dimensional phase space are given by
\begin{eqnarray}
    \left\{\theta, p\right\} = \frac{\gamma \kappa}{6} \,, \quad \left\{\lambda, p_{\lambda}\right\} =\left\{\mu, p_{\mu}\right\} =1,\quad \{\phi^A,\pi_B\}=\delta^A_B,\quad \{N,p_N\}=1,
\end{eqnarray}
where $\gamma$ denotes the Barbero-Immirzi parameter and $\kappa=8\pi G$ with $G$ being Newton's constant.
The model we consider has a primary Hamiltonian of the form
\begin{eqnarray}
 H^{P} = NH_0(\theta,p,\mu,\varphi^A,\pi_A)+ \lambda (\mu - f(p,\theta)) + \Lambda_{\lambda} p_{\lambda} + \Lambda_{\mu} p_{\mu}+\Lambda_Np_N,
\end{eqnarray}
where $\Lambda_{\lambda},\Lambda_{\mu},\Lambda_N$ are Lagrange multipliers and $H_0(\theta,p,\mu,\varphi^A,\pi_A)$ takes the following form 
\begin{equation}
   H_0(\theta,p,\mu,\varphi^A,\pi_A) = %
   -\frac{6 \sqrt{p} h( \mu \theta)}{\gamma ^2 \kappa \mu^2} 
   +H_{0}^{\rm matter}(\phi^A,\pi_A,p) .
\end{equation}
We consider a polymerization of the connection variable $\theta$ in the gravitational sector with a generic function $h(\mu \theta)$ satisfying
\begin{eqnarray}
    \lim_{\mu \to 0} \frac{h(\mu \theta)}{\mu^2} = \theta^2.
\end{eqnarray}
 This provides the possibility to either consider this model as a classical starting point for a later loop quantization or to stay at the classical level and consider it as an effective model. Choosing this function to be $h(\mu \theta) = \sin(\mu \theta)^2$ gives the usual LQC effective Hamiltonian. The function $f(\theta,p)$ is an up to now arbitrary function and can be later chosen to have a specific form in order to reproduce the wanted $\mubar$-scheme. The system has three primary constraints 
\begin{eqnarray}
    C_{p_{\lambda}} = p_{\lambda} \approx 0 \,,  \qquad  C_{p_{\mu}} = p_{\mu} \approx 0 
    \,,  \qquad  C_{p_{N}} = p_{N} \approx 0 .
\end{eqnarray}
The stability of $C_{p_{\lambda}}$ leads to the following gauge fixing condition for $\mu$,
\begin{eqnarray}
\label{eq:Cmu}
C_{\mu} &=& \dot{ C}_{p_{\lambda}} = \mu - f_\mu(\theta,p) \approx 0 \,
\end{eqnarray}
The stability of $C_{p_{\mu}}$ gives
\begin{eqnarray}
    C_{\lambda} = \dot{C}_{p_{\mu}}= \lambda+\frac{12 \sqrt{p} h ( \mu \theta  )}{\gamma ^2 \kappa  \mu ^3} - \frac{6 \theta  \sqrt{p} h'(  \mu \theta  )}{\gamma ^2 \kappa  \mu ^2}=:\lambda + g(\theta,p,\mu) \approx 0 \,,
\end{eqnarray}
with $h'(X) = \partial_X h(X)|_{X = \mu \theta}$. Note that the matter contribution $H_0^{\rm matter}$ does not contribute to $C_\lambda$ because it does not depend on $\mu$ as long as we do not consider a polymerization of the matter part. Furthermore, the requirement of the stability for $C_{p_{N}}$ yields the Hamiltonian constraint 
\begin{eqnarray}
\label{eq:HamConCosmo}
    C_0 = H_0(\theta,p,\mu,\varphi^A,\pi_A) \approx 0 .
\end{eqnarray}
The stability of $C_{\mu}$ and $C_{\lambda}$ fix Lagrangian multiplier $\Lambda_{\mu}$ and $\Lambda_{\lambda}$ respectively, whereas $C_0$ is already stable. 
~\\
The canonical transformation that we want to construct is a so-called partial Kucha\v{r} decomposition of the original phase space. We call it partial here because we will perform such a decomposition only with respect to part of the constraints, namely those related to the choice of the $\mubar$-scheme. Strictly speaking, we should also call the Dirac observable partial Dirac observables because they span the partially reduced phase space but not the fully reduced one which needs to  involve observables with respect to the Hamiltonian constraints. In the following, we will still call the partial Dirac  observables just Dirac observables but have in mind that the Hamiltonian constraint has not been considered yet but could easily be included following for instance the strategy used in \cite{Giesel:2020raf} by using an additional observable map based on matter clocks and where details on the construction of such observable maps are presented.

For this purpose we will use the observable map as well as its dual version introduced in \cite{Fahn:2022zql} where a Kucha\v{r} decomposition \cite{Kuchar:1972,Hajicek:1999ht} of the phase space was constructed in the context of linearized gravity. This is necessary because originally we want to choose some of the involved constraints as new canonical coordinates and this requires that each constraint commutes with all but one. Because of this we cannot use the original set of constraints for this purpose. Applying different observable maps successively will, however, yield new phase space coordinates that can be used for this purpose. We will then show that the application of the different observables maps is indeed a canonical transformation on the extended phase space. 
~\\
~\\
As a first step, we consider the two first class constraints $C_{p_\mu}$ and $C_{p_\lambda}$. Because there are elementary momentum variables in phase space they mutually commute with all remaining phase space variables except $\mu,\lambda$ and as a consequence $C_\mu,C_\lambda$. In particular $\theta$ and $p$ are obviously Dirac observables with respect to $C_{p_\mu}, C_{p_\lambda}$. This means that the corresponding observable maps are defined as
\begin{equation}
\label{eq:ObsmapComso}
 {\cal O}^{p_\mu}_f:=\sum\limits_{n=0}^\infty \frac{(-1)^n(C_\mu)^n}{n!} \{f,C_{p_{\mu}}\}_{(n)},
\quad
{\cal O}^{p_\lambda}_f:=\sum\limits_{n=0}^\infty \frac{(-1)^n(C_\lambda)^n}{n!} \{f,C_{p_{\lambda}}\}_{(n)},
\end{equation}
where $\{.,.\}_{(n)}$ denotes the nested Poisson bracket with $\{f,g\}_{(0)}=f$ and $\{f,g\}_{(n+1)}=\{\{f,g\}_{(n)},g\}$
and $f$ is a function on phase space that becomes the identity map for functions $f$ that depend on $(\theta,p)$ only. In the language of the relational formalism the constraints $C_\mu,C_\lambda$ play the role of clocks. Now given the observable map in \eqref{eq:ObsmapComso} we can define its dual map along the lines of the work in \cite{Fahn:2022zql} where the role of clocks and first class constraints is interchanged. However, if we want to consider dual observable maps for both pairs $(C_\mu,C_{p_\mu})$ and $(C_\lambda,C_{p_\lambda})$ we need that the 'clocks' $C_\mu,C_\lambda$ commute at least weakly. Considering the Dirac matrix in \eqref{eq:FullDiracMatrixCosmo} we see that this is not the case. Since we aim at constructing a canonical transformation we even need the stronger requirement that the clocks strongly commute. This can be achieved by successively applying the dual observable map as follows: in the first step we consider the following set of variables
\begin{eqnarray}
 Q^1:= C_\lambda,\quad P_1:=C_{p_\lambda},\quad  Q^2:&=&{\cal O}_{C_\mu}^{C_\lambda},\quad  P_2:={\cal O}_{C_{p_\mu}}^{C_\lambda} ,\quad 
 \end{eqnarray}
 with the dual observable map
 \begin{equation}
\label{eq:ObsmapDualClambda}
 {\cal O}^{C_\lambda}_f:=\sum\limits_{n=0}^\infty \frac{(C_{p_\lambda})^n}{n!} \{f,C_{\lambda}\}_{(n)}.
\end{equation}
By construction we have $\{{\cal O}^{C_\lambda}_f,C_\lambda\}=0$. Using the result in \cite{Thiemann:2004wk} on the  Poisson algebra of the so constructed Dirac observables and carrying it over to the dual observable map we obtain
\begin{equation}
 \{{\cal O}_{C_\mu}^{C_\lambda} ,  {\cal O}_{p_\mu}^{C_\lambda}\}
 = {\cal O}_{\{C_\mu,p_\mu\}^{C_\lambda}_D} = {\cal O}_{\{C_\mu,p_\mu\}}=1,
\end{equation}
where ${\{.,.\}^{C_\lambda}_D}$ denotes the submatrix of the Dirac bracket in \eqref{eq:FullDiracMatrixCosmo} that includes the Poisson brackets of the pair $(C_\lambda,C_{p_\lambda})$ only. Given this we already have $\{Q^j,P_k\}=\delta^j_k$ with $j,k=1,2$. To extend that set by variables corresponding to the gravitational and matter degrees of freedom that have the property to mutually commute with $Q^j,P_j$ for $j=1,2$ and satisfy standard CCR we define
\begin{equation}
 Q^3:={\cal O}^{C_\lambda, {\cal O}^{C_\lambda}_{C_\mu}}_{\theta},\quad  
 P_3:={\cal O}^{C_\lambda, {\cal O}^{C_\lambda}_{C_\mu}}_{p},\quad 
 Q^4:=N, \quad P_4:=p_N,\quad 
 Q^I:={\cal O}^{C_\lambda, {\cal O}^{C_\lambda}_{C_\mu}}_{\phi^A},\quad 
 P_I:={\cal O}^{C_\lambda, {\cal O}^{C_\lambda}_{C_\mu}}_{\pi_A},
\end{equation}
with $I\in 5,\cdots, L_{\#}+4$ being an index set covering the matter degrees of freedom for all $A$ at the level of the observables.  The notation  ${\cal O}_f^{C_\lambda, {\cal O}^{C_\lambda}_{C_{p_{\mu}}}}$ means that in the first step we apply the dual observable map from \eqref{eq:ObsmapDualClambda} on $f$ and afterwards the following dual observable map
\begin{equation}
 \label{eq:DualObsmapCmu}
 {\cal O}^{{\cal O}^{C_\lambda}_{C_\mu}}_f:=\sum\limits_{n=0}^\infty \frac{({\cal O}^{C_\lambda}_{C_{p_\mu}})^n}{n!} \{f,{\cal O}^{C_\lambda}_{C_{\mu}}\}_{(n)}. 
\end{equation}
For the canonical pair $(N,p_N)$ we have used that both variables trivially commute with $C_\lambda,C_\mu$ and thus these maps act like the identity map on $N,p_N$.
~\\
Now it remains to show that $Q^3,P_3,Q^4,P_4,Q^I,P_I$ satisfy standard CCR and have vanishing Poisson brackets with $Q^1,Q^2,P_1,P_2$. For the pair $(Q^4,P_4)$ this is the case  because we have $\{Q^4,P_4\}=1$ and none of the constructed observables depends on $N,p_N$ so that they mutually commute with them. To show it also for the remaining set $Q^3,P_3,Q^I,P_I$ a little more work is needed. By construction, $C_\lambda$ and ${\cal O}^{C_\lambda}_{C_\mu}$ commute and hence the order of the application of the two (dual) observable maps is irrelevant and will yield a quantity that commutes with $Q^1$ as well as $Q^2$ by construction.  To compute the Poisson bracket with the momenta we use
\begin{eqnarray}
\label{eq:DoubleObs}
 {\cal O}^{C_\lambda, {\cal O}^{C_\lambda}_{C_\mu}}_{f} &=&
 \sum\limits_{n=0}^\infty \frac{1}{n!}\left({\cal O}^{C_\lambda}_{p_\mu}\right)^n \{{\cal O}^{C_\lambda}_{f}, {\cal O}^{C_\lambda}_{C_\mu}\}_{(n)}  = \sum\limits_{n=0}^\infty \frac{1}{n!}\left({\cal O}^{C_\lambda}_{p_\mu}\right)^n 
 {\cal O}^{C_\lambda}_{\{f,C_\mu\}_{D(n)}^{C_\lambda}},
\end{eqnarray}
where $\{f,g\}_{D(n)}^{C_\lambda}$ denotes the iterated Dirac bracket with respect to the pair $(C_\lambda,C_{p_\lambda})$ with $\{f,g\}_{D(0)}^{C_\lambda}=f
$ and $\{f,g\}_{D(n+1)}^{C_\lambda}=\{\{f,g\}_{D(n)}^{C_\lambda},g\}_D^{C_\lambda}$.

The last step in \eqref{eq:DoubleObs} can be easily shown using again the result on the observable algebra from \cite{Thiemann:2004wk} iteratively as well as a proof by induction, which we present in the lemma 1 below. The lemma 1 is not discussed in \cite{Thiemann:2004wk} but can be easily proven using the theorem presented in \cite{Thiemann:2004wk} and we present the proof in appendix \ref{app:Lemma}.
\begin{customlemma}{1}
\label{lemma1}
For the iterated Poisson bracket of observables we have
\begin{equation}
\{ {\cal O}^{C_\lambda}_f, {\cal O}^{C_\lambda}_g \}_{(n)} = {\cal O}^{C_\lambda}_{\{f,g\}_{D(n)}^{C_\lambda}}.
\end{equation}
\end{customlemma}
~\\
Using these results the Poisson brackets with the momenta yield
\begin{eqnarray*}
 \{Q^3,P_2\} &=& \{{\cal O}^{C_\lambda, {\cal O}^{C_\lambda}_{C_\mu}}_{\theta} , {\cal O}_{C_{p_\mu}}^{C_\lambda} \} 
 =\sum\limits_{n=0}^\infty\frac{1}{n!} \{({\cal O}^{C_\lambda}_{C_{p_\mu}})^n {\cal O}^{C_\lambda}_{\{\theta,C_\mu\}_{D(n)}^{C_\lambda}} , {\cal O}_{C_{p_\mu}}^{C_\lambda} \} \\
 &=&\sum\limits_{n=0}^\infty\frac{1}{n!} ({\cal O}^{C_\lambda}_{C_{p_\mu}})^{n} \{{\cal O}^{C_\lambda}_{\{\theta,C_{\mu}\}_{D(n)}^{C_\lambda}},  {\cal O}_{C_{p_\mu}}^{C_\lambda} \} 
 =\sum\limits_{n=0}^\infty\frac{1}{n!} ({\cal O}^{C_\lambda}_{C_{p_\mu}})^{n} {\cal O}^{C_\lambda}_{\{\{\theta,C_{\mu}\}_{D(n)}^{C_\lambda},C_{p_\mu}\}_D^{C_\lambda}},
\end{eqnarray*}
where we used the result of lemma 1 in the second and one before the last step. 
~\\
Now we consider the Dirac bracket $\{\theta,C_{\mu}\}_{D(n)}^{C_\lambda}$. For $n=0$ we have $\{\theta,C_{\mu}\}_{D(0)}^{C_\lambda}=\theta$. For $n\geq 1$ we get
\begin{equation*}
\{\theta,C_{\mu}\}_{D(n)}^{C_\lambda}=\{\theta,C_{\mu}\}_{(n)}=\{\theta,f_{\mu}\}_{(n)}, 
\end{equation*}
where by assumption the function $f_\mu$ depends on the gravitational degrees of freedom only and thus will $\{\theta,f_{\mu}\}_{(n)}$ do so  as well. Here we used in the first step that $\theta,C_\mu$ both commute with $p_\lambda$. Since any partial derivative of $f$ as well as $\theta$ commutes with $p_\lambda$ we obtain for $n\geq 1$
\begin{equation*}
 \{\{\theta,C_{\mu}\}_{D(n)}^{C_\lambda},C_{p_\mu}\}_D^{C_\lambda}
 = \{\{\theta,C_{\mu}\}_{D(n)}^{C_\lambda},C_{p_\mu}\}=\frac{\partial}{\partial\mu}\{\theta,f_{\mu}\}_{(n)}=0.
\end{equation*}
For $n=0$ the nested Dirac bracket also vanishes because $\{\theta,C_{p_\mu}\}=0$. From these results we immediately end up with
\begin{equation}
 \{Q^3,P_2\}=\sum\limits_{n=0}^\infty\frac{1}{n!} ({\cal O}^{C_\lambda}_{C_{p_\mu}})^{n} {\cal O}^{C_\lambda}_{\{\{\theta,C_{\mu}\}_{D(n)}^{C_\lambda},C_{p_\mu}\}_D^{C_\lambda}}
 =\sum\limits_{n=0}^\infty\frac{1}{n!} ({\cal O}^{C_\lambda}_{C_{p_\mu}})^{n} {\cal O}^{C_\lambda}_{0}=0.
\end{equation}
For the momentum $P_1$ the calculation of the Poisson bracket leads to
\begin{eqnarray*}
 \{Q^3,P_1\} &=& \{{\cal O}^{C_\lambda, {\cal O}^{C_\lambda}_{C_\mu}}_{\theta} , C_{p_\lambda} \} 
 =\{{\cal O}^{C_\lambda, {\cal O}^{C_\lambda}_{C_\mu}}_{\theta} ,C_{p_\lambda} \} 
 =\frac{\partial}{\partial \lambda}\sum\limits_{n=0}^\infty \frac{({\cal O}^{C_\lambda}_{C_{p_\mu}})^n}{n!} 
\{{\cal  O}^{C_\lambda}_\theta,{\cal O}^{C_\lambda}_{C_{\mu}}\}_{(n)}=0.
\end{eqnarray*}
The last step follows from the fact that if we consider \eqref{eq:DoubleObs} we see that for functions $f$ that do not depend on $p_\lambda$ the term linear in $\lambda$ in $C_\lambda$ does not contribute to the nested Poisson bracket. Hence, neither ${\cal O}^{C_\lambda}_{C_{p_\mu}}$ nor ${\cal O}^{C_\lambda}_{C_\mu}$ depend on $\lambda$. Since also $\partial_\lambda{\cal O}^{C_\lambda}_\theta=0$ the nested Poisson bracket in the above equation is independent of $\lambda$ too. 
~\\
The same steps can be performed if we replace $Q^3$ by $P_3$ as well as the matter degrees of freedom $Q^I,P_I$ so that we have shown that indeed $Q^3,P_3,Q^I,P_I$ mutually commute with $Q^1,Q^2,P_1,P_2$.

~\\
Finally we still need to show that $Q^3,P_3$ as well as $Q^I,P_I$ satisfy standard CCR and mutually commute. For showing that they satisfy standard CCR we restrict our discussion to the canonical pair ${Q^3,P_3}$ and discuss the matter variables afterwards.  For this purpose it is more convenient to directly work with the second class constraints $(C_\lambda, {\cal O}^{\rm C_\lambda}_{C_\mu}, C_{p_\lambda}, {\cal O}^{C_\lambda}_{p_\mu})$ and the variables ${\cal O}^{C_
\lambda}_\theta, {\cal O}^{C_\lambda}_p$ and their corresponding algebra.  Then we obtain
\begin{eqnarray}
\label{eq:CCRGeoObs}
\{{\cal O}^{C_\lambda, {\cal O}^{C_\lambda}_{C_\mu}}_{\theta} , {\cal O}^{C_\lambda, {\cal O}^{C_\lambda}_{C_\mu}}_{p} \} &=&  {\cal O}_{\{{\cal O}^{C_\lambda}_\theta, {\cal O}^{C_\lambda}_p\}_{D}^{C_\mu}}= {\cal O}_{\{{\cal O}^{C_\lambda}_\theta, {\cal O}^{C_\lambda}_p\}}=\frac{\gamma\kappa}{6}
\end{eqnarray}
with 
\begin{eqnarray*}
\{{\cal O}^{C_\lambda}_\theta, {\cal O}^{C_\lambda}_p\}_{D}^{C_\mu}
&: =&\{{\cal O}^{C_\lambda}_\theta, {\cal O}^{C_\lambda}_p\} + \{{\cal O}^{C_\lambda}_\theta, {\cal O}^{C_\lambda}_{C_\mu}\} \{{
\cal O}^{C_\lambda}_{C_{p_\mu}}, {\cal O}^{C_\lambda}_p\} 
- \{{\cal O}^{C_\lambda}_\theta, {
\cal O}^{C_\lambda}_{C_{p_\mu}}\}\{ {\cal O}^{C_\lambda}_{C_\mu}, {\cal O}^{C_\lambda}_p\} \\
&=&\{{\cal O}^{C_\lambda}_\theta, {\cal O}^{C_\lambda}_p\} =\frac{\gamma\kappa}{6}.
\end{eqnarray*}
Here we used that 
\begin{eqnarray*}
 \{{\cal O}^{C_\lambda}_\theta, {\cal O}^{C_\lambda}_p\} 
 &=& {\cal O}_{\{\theta,p\}_D^{C_\lambda}} = {\cal O}_{\{\theta,p\}} =\frac{\gamma\kappa}{6}, \\
 \{{
\cal O}^{C_\lambda}_{C_{p_\mu}}, {\cal O}^{C_\lambda}_p\} &=& 
{\cal O}_{\{C_{p_\mu},p\}_D^{C_\lambda}} = {\cal O}_{\{C_{p_\mu},p\}} =0 \\
\{{\cal O}^{C_\lambda}_\theta,{\cal O}^{C_\lambda}_{C_{p_\mu}}\} &=& 
{\cal O}_{\{\theta , C_{p_\mu}\}_D^{C_\lambda}} = {\cal O}_{\{\theta, C_{p_\mu}\}} =0,
 \end{eqnarray*}
 where in all three cases the Dirac bracket agrees with the Poisson bracket because $\theta,p$ commute with $C_{p_\lambda}$ and we further use that $\theta,p$ also Poisson commute with $C_{p_\mu}$. The result in \eqref{eq:CCRGeoObs} can be directly carried over to $Q^I,P_I$ the observables corresponding to the elementary matter phase space variables $\phi^A,\pi_A$ because the crucial property needed for this result is that both $\theta,p$ Poisson commute with $C_{p_\lambda}$ and $C_{p_\mu}$ which is also given for all $
\phi^A,\pi_A$. Moreover, this property is also sufficient to show that the sets $Q^3,P_3$ $Q^I,P_I$ mutually commute because also in this computation the relevant Dirac brackets reduce to their corresponding Poisson brackets and then we can use that the original phase space variables $p,\theta$ and $\phi^A,\pi_A$ are sets of mutually commuting elementary variables in phase space. 

 ~\\ 
 Summarizing the canonical transformation that allows to perform a (partial) Kucha\v{r} decomposition of the $(8+2L_{\#})$ dimensional phase space separating the physical degrees of freedom from the gauge degrees of freedom is given by
 \begin{eqnarray*}
Q^1&:=& C_\lambda,\quad P_1:=C_{p_\lambda},\quad  Q^2:= {\cal O}_{C_\mu}^{C_\lambda},\quad P_2:={\cal O}_{C_{p_\mu}}^{C_\lambda}, \quad 
 Q^3:={\cal O}^{C_\lambda, {\cal O}^{C_\lambda}_{C_\mu}}_{\theta},\quad  
 P_3:={\cal O}^{C_\lambda, {\cal O}^{C_\lambda}_{C_\mu}}_{p},\\
 Q^4&:=&N,\quad P_4:=p_N,\quad 
 Q^I:={\cal O}^{C_\lambda, {\cal O}^{C_\lambda}_{C_\mu}}_{\phi^A},\quad 
 P_I:={\cal O}^{C_\lambda, {\cal O}^{C_\lambda}_{C_\mu}}_{\pi_A},
 \end{eqnarray*}
 with $I=5,\cdots, L_{\#}+4$.
 In terms of the original variables $\theta,p,\mu,p_\mu,\lambda,p_\lambda,\phi^A,\pi_A$ these are rather complicated functions consisting of partly nested power series with phase space depended coefficients. However, as shown above the new variables satisfy standard CCR
 \begin{eqnarray*}
     \{Q^J,P_K\} &=& \widetilde{\delta}^J_K,\quad \{Q^J,Q^K\} = 0, \quad \{P_J,P_K\} =0,\quad J,K=1,\dots, L_{\#}+4
 \end{eqnarray*}
with 
\begin{eqnarray*}
 \widetilde{\delta}^J_K=\delta^J_K \quad {\rm for}\quad J,K=1,\dots, L_{\#}+4 {\quad} {\rm if}\quad J=K\not= 3 \quad {\rm and}\quad
 \widetilde{\delta}^3_3=\frac{\gamma\kappa}{6}.
\end{eqnarray*}
 The primary Hamiltonian in the new variables expressed as a function in terms of the old canonical variables looks similarly complicated but has in the new variables the following form:
 \begin{eqnarray}
 \label{eq:PrimHamObs}
 H^{P} &=& H_0(Q^3,P_3,Q^1+f_\mu(Q^3,P^3),Q^I,P_I)+ (Q^2-g(Q^3,P_3,Q^1+f_\mu(Q^3,P_3)))Q^1   \nonumber \\
&& + \Lambda_{\lambda} P_2 + \Lambda_{\mu} P_1. \nonumber
\end{eqnarray}
If we consider the Hamiltonian constraint $C_0$ in \eqref{eq:HamConCosmo} on the partially reduced phase space with respect to the constraints encoded in $Q^1,P_1,Q^2,P_2$ that we obtain by setting $Q^1=Q^2=P_1=P_2=0$ we end up with
\begin{eqnarray*}
C_0 &=& H_0(Q^3,P_3,f_\mu(Q^3,P^3),Q^I,P_I)
\end{eqnarray*}
showing that we obtain the Hamiltonian constraint in the $\mubar$-scheme characterized by the choice of the function $f_\mu$ in \eqref{eq:Cmu} 
and in this sense dynamically implemented the $\mubar$ by a canonical transformation on an extended phase space in terms of a Kucha\v{r}-decomposition and the corresponding partial reduction afterwards. This can now be taken as the starting point to either implement a Dirac quantization or reduced quantization with respect to the Hamiltonian constraint and the remaining primary constraint $p_N$ which are both first class or work with corresponding effective models.
\subsection{Gauge-fixed model for cosmology}
\label{sec:GFmodel}
For the corresponding gauge fixed version of the model introduced in section \ref{sec:DiracObsCanTrafo} we use that 
the set of constraints $C =\left\{C_{\lambda},C_{\mu},C_{p_{\lambda}},C_{p_{\mu}}\right\}$ form a second class system. The Dirac matrix has the form
\begin{eqnarray}
\label{eq:FullDiracMatrixCosmo}
    M = \left(
        \begin{array}{cccc}
        0 & A_{\lambda \mu } & 1  & A_{\lambda p_{\mu}} \\
        -A_{\lambda \mu } & 0 & 0 & 1 \\
        -1  & 0 & 0 & 0\\
        -A_{\lambda p_{\mu}} & -1 & 0 & 0 \\
        \end{array}
    \right) \,, \quad M^{-1} = \left(
        \begin{array}{cccc}
         0 & 0 & -1 & 0 \\
        0 & 0 & A_{\lambda p_{\mu}} & -1\\
         1 & -A_{\lambda p_{\mu}} & 0 & A_{\lambda \mu } \\
         0 & 1 & -A_{\lambda \mu } & 0 \\
        \end{array}
        \right)
\end{eqnarray}
with
\begin{eqnarray}
    A_{\lambda p_{\mu}} &=& - \frac{6 \sqrt{p} \left(6 h(\theta  \mu )+\theta  \mu  (-4 h'(\theta  \mu )+\theta  \mu h''( \theta  \mu ))\right)}{\gamma ^2 \kappa  \mu ^4}\\%
    A_{\lambda \mu } &=& \frac{\partial_\theta f_\mu (2 h (\theta  \mu ) - \theta  \mu h'( \theta  \mu ))+2 \mu  p \partial_p f_\mu (-h'( \theta  \mu ) + \theta  \mu  h''( \theta  \mu ))}{2 \gamma  \mu ^3 \sqrt{p}}.
\end{eqnarray}
The Poisson brackets between $\theta,p$ and constraints $C$ are non-trivial for $C_{\lambda,\mu}$ only. As a result, the Dirac brackets between $\theta$ and $p$ recovers their Poisson brackets for generic gauge fixing function $f$, namely
\begin{eqnarray}
\label{eq:DBCosmograv}
    \left\{\theta, p\right\}_D = \frac{\gamma \kappa}{6}\,, \quad \left\{\theta, \theta \right\}_D = \left\{p, p\right\}_D =0
\end{eqnarray}
Likewise we obtain for the matter degrees of freedom
\begin{eqnarray}
\label{eq:DBCosmomatter}
    \left\{\phi^A, \pi_B\right\}_D = \delta^A_B\,, \quad \left\{\phi_A, \phi_B \right\}_D = \left\{\pi_A, \pi_B\right\}_D =0,\quad A,B\in 1,\dots, L_{\#}.
\end{eqnarray}
The remaining Hamiltonian constraint has in the partially reduced phase space the following form
\begin{equation}
 C = H_0(\theta,p,f_\mu(\theta,p),\phi^A,\pi_A).   
\end{equation}
The canonical transformation in section \ref{sec:DiracObsCanTrafo} and the gauge fixing here does not further restrict the function $f_\mu(\theta,p)$ and particularly allows to choose the special case of a constant function which corresponds exactly to the $\mu_0$-scheme in LQC. 
~\\
Now if we require in addition invariance under the following transformation
\begin{equation}
\theta \rightarrow \alpha\theta,\quad p \rightarrow \alpha^2 p    
\end{equation}
then this further restricts the form of $f_\mu(\theta,p)$ to $f_\mu(p)=\frac{c_1}{\sqrt{|p|}}$ where $c_1$ is some constant, in the case of the cosmological models this is the only possible scale invariant choice. Choosing this constant to be $c_1:=\sqrt{\Delta}$ with $\Delta:=2\sqrt{3}\pi\gamma\ell_p^2$, where $\gamma$ is the Barbero-Immirzi parameter and $\ell_p$ denotes the Planck length, we end up with the usual $\mubar$ scheme used in LQC \cite{Ashtekar:2006wn}. 
~\\

In this section we have shown that we can implement the choice of a generic phase space dependent function in the argument of the polymerization function dynamically in the following sense: In section \ref{sec:DiracObsCanTrafo} this was achieved in terms of a canonical transformation on the extended phase space that allows a Kucha\v{r} decomposition \cite{Kuchar:1972,Hajicek:1999ht} of the extended phase space. We extended the usual kinematical phase space used in LQC by two additional canonical pairs which come along with two first class constraints so that the number of physical degrees of freedom does not change.  A later reduction to the physical sector associated with these additional first class constraints then corresponds to a specific choice of a $\mubar$-scheme. Furthermore, we showed that in this extended phase space we can also understand the choice of a given $\mubar$-scheme as a gauge fixing.  We will extend our analysis to the spherically symmetric case in the next section.

\section{Implementing the $\mubar$-scheme in spherically symmetric models}\label{sec:BH}
In this section we want to generalize the formalism used for cosmology in \ref{sec:DiracObsCanTrafo} to spherically symmetric models. The main difference is that on the one hand we deal with a $1+1$-dimensional field theory and thus the diffeomorphism constraint is no longer trivial. Moreover, since the number of gravitational (field) variables is larger here we need to extend the phase space by more variables in order to be able to allow an independent polymerization for the different connection variables involved. In case we describe the Schwarzschild black hole using isometry with Kantowski-Sachs vacuum cosmology, where the diffeomorphism is trivially satisfied, the system is similar to cosmological models with a larger phase space.
\subsection{Extended phase space for spherically symmetric models: Canonical Transformation and Dirac Observables}
\label{sec:DiracObsCanTrafoBH}
In spherically symmetric symmetry reduced case after implementing the Gau\ss{} constraint the Ashtekar-Barbero variables $(A^j_a,E^a_j)$ have the following form
\begin{eqnarray*}
A_a^j \tau_j \mathrm{~d} X^a & = & 2\gamma K_x(x) \tau_1 \mathrm{~d} x+\left(\gamma K_\varphi(x) \tau_2+\frac{\partial_x E^x(x)}{2E^\varphi(x)} \tau_3\right) \mathrm{d} \theta \\
&&+\left(\gamma K_\varphi(x) \tau_3-\frac{\partial_x E^x(x)}{2E^\varphi(x)} \tau_2\right) \sin (\theta) \mathrm{d} \varphi+\cos (\theta) \tau_1 \mathrm{~d} \varphi \\
E^a_j \tau^j \frac{\partial}{\partial X^a} & = &E^x(x) \sin (\theta) \tau_1 \partial_x+\left(E^\varphi(x) \tau_2\right) \sin (\theta) \partial_\theta+\left(E^\varphi(x) \tau_3\right) \partial_{\varphi},  
\end{eqnarray*}
where $X^a=(x,\theta,\varphi)$, $a=1,2,3$ denote spherical coordinates, as before, $\gamma$ the Barbero-Immirzi parameter and $\tau_j=-\frac{1}{2}\sigma_j$ with $\sigma_j$ being the Pauli matrices. The with respect to the Gau\ss{} constraint partially gauge fixed phase space includes two canonical pairs denoted by $\left(K_x(x), E^x(x)\right)$ and $\left(K_{\varphi}(x), E^{\varphi}(x)\right)$. In addition, as in the cosmological case we consider a set of matter degrees of freedom that we do not further specify whose phase space variables are $(\phi^A(x),\pi_A(x))$ with $A\in 1,\dots,L_{\#}$. Their non-vanishing Poisson brackets read
\begin{equation*}
 \{K_x(x), E^x(y)\} = G \delta(x,y)\quad  \{K_\varphi(x), E^\varphi(y)\} =G \delta(x,y)\quad \{\phi^A(x),\pi_B(y)\}=\delta^A_B\delta(x,y),
\end{equation*}
with $G$ being Newton's constant. We consider this phase space as the starting point for our extension. We introduce the following additional set of canonical pairs
\begin{eqnarray*}
 (e_1(x),p_{e_1}(x)),\quad (e_2(x),p_{e_2}(x)),\quad (\mu_x(x),p_{\mu_x}(x)),\quad (\mu_\varphi(x),p_{\mu_\varphi}(x)).  
\end{eqnarray*}
with standard CCR
\begin{eqnarray*}
\{ e_I(x),p_{e_J}(y)\}=\delta^I_J \delta(x,y),\quad I,J=1,2\quad 
\{\mu_K(x),p_{\mu_M}(y)\}=\delta^K_L \delta(x,y),\quad K,L=x,\varphi .
\end{eqnarray*}
Note that here we considered already the partially reduced phase space with respect to the constraint $p_N(x)$ associated with the momentum of the lapse function $N(x)$ such that the latter becomes a Lagrange multiplier here and is not part of the elementary phase space variables. We start with the following primary Hamiltonian
\begin{eqnarray}\label{PH-sph}
     H &=& \int \text{d}x\, \Big[ \mathcal{C}^{\Delta}_{\rm tot} + N^x \mathcal{C}^{\rm tot}_x + e_1 (\mu_x - f_{\mu_x}(E^x,E^{\varphi},K_x,K_{\varphi})) + e_2 (\mu_{\varphi} - f_{\mu_{\varphi}}(E^x,E^{\varphi},K_x,K_{\varphi})) \nonumber\\
     &&\qquad \qquad \qquad+ \Lambda_{N^x} p_{N^x} + \Lambda_{e_1} p_{e_1} + \Lambda_{e_2} p_{e_2} + \Lambda_{\mu_x} p_{\mu_{x}} + \Lambda_{\mu_{\varphi}} p_{\mu_{\varphi}} \Big] (x),
\end{eqnarray}
where we considered a partial gauge fixing that yields for the lapse function $N=1$. This can for instance be achieved by choosing  a suitable external matter field that is chosen as a temporal reference field such that $N=1$ is implemented.
As a consequence,  in contrast to the cosmological model in \ref{sec:DiracObsCanTrafo} here the polymerized contribution $\mathcal{C}^{\Delta}_{\rm tot}$ is not a secondary constraint but plays the role of a physical Hamiltonian instead, once the reduction with respect to the spatial diffeomorphism constraint has also been performed. We consider here a partial gauge fixing because, in general, we also expect further restrictions on the possible polymerizations from the closure of the algebra of constraints. Such a more general analysis is beyond the scope of this article and is planned for future work, although the analysis performed here will be useful as a first step.
~\\
Here $\mathcal{C}^{\Delta}_{\rm tot}$  and $\mathcal{C}^{\rm tot}_x$ are given by
\begin{equation}
 \mathcal{C}^{\Delta}_{\rm tot} = \mathcal{C}^{\Delta} + \mathcal{C}^{\rm matter}(E^x,E^\varphi,\phi^A,\pi_A) \quad
 \mathcal{C}^{\rm tot}_x = \mathcal{C}_x + \mathcal{C}_x^{\rm matter}(\phi^A,\pi_A)
\end{equation}\
with
\begin{eqnarray}
    \mathcal{C}^{\Delta}(x)&=&\frac{1}{2 G}\left[\frac{{{E^{\varphi}}}}{\sqrt{{{E^x}}}}\qty( \frac{f_1(\mu_{x} K_{x},\ \mu_{\varphi} K_{\varphi})}{\mu_{\varphi}^2} + \Gamma^2 - 1  - 4 E^x\frac{ f_2(\mu_{x} K_{x},\ \mu_{\varphi} K_{\varphi})}{\mu_{x} \mu_{\varphi} E^{\varphi}} )+2\sqrt{E^x} \Gamma'
\right](x)\\
    \mathcal{C}_x &=& \frac{1}{G}\left(E^{\varphi} {K_{\varphi}} ' - K_x {E^x}'\right)(x)
\end{eqnarray}
with $\Gamma(x) =\qty(\frac{  {{E^x}}'}{2{{E^{\varphi}}} })(x)$ the spin connection, and $f_1,f_2$ represent general polymerization functions satisfying
\begin{eqnarray}\label{eq:poly_f1f2_BH}
    \lim_{\mu_{\varphi} \to 0} \frac{ f_1(\mu_{x} K_{x}, \ \mu_{\varphi} K_{\varphi})}{\mu_{\varphi}^2} = K_{\varphi}^2 \,, \qquad \lim_{\mu_x,\mu_{\varphi} \to 0} \frac{ f_2(\mu_{x} K_{x},\ \mu_{\varphi} K_{\varphi})}{\mu_{x} \mu_{\varphi}} = K_x K_{\varphi} 
\end{eqnarray}
In order that $\mathcal{C}^{\Delta}$ is a scalar density of weight $1$, we are forced to choose $\mu_x$ being as scalar with density weight $-1$ since $K_x$ has density weight $1$. Furthermore,  $\mu_{\varphi}$ needs to have density weight $0$. As a result, the only allowed  functions $ f_{\mu_x}$ and $f_{\mu_{\varphi}}$ will be of the form 
\begin{eqnarray}
\label{eq:Choicefmuxfmuphi}
     &f_{\mu_x}(E^x,E^{\varphi},K_x,K_{\varphi}) = (E^{\varphi})^{-1}  \tilde{f}_{\mu_x}(E^x,K_{\varphi},(E^{\varphi})^{-1}K_{x})\, , \\
     \label{eq:Choicefmuxfmuphi1}
     &f_{\mu_{\varphi}}(E^x,E^{\varphi},K_x,K_{\varphi}) =  \tilde{f}_{\mu_{\varphi}}(E^x,K_{\varphi},(E^{\varphi})^{-1}K_{x}).
\end{eqnarray}
Here the appearance of $E^{\varphi}$, and the combination of $(E^{\varphi})^{-1}K_{x}$ are essential in order to have density weight of one and zero for $f_{\mu_x}$ and $f_{\mu_{\varphi}}$ respectively. This completely removes the possibility of the usual $\mu_0$ scheme in the effective dynamics if we have polymerization of $K_x$.
If we require in addition the scaling invariance of $f_1$ and $f_2$ and remove the $K$ dependence in the gauge fixing function, the only possible ansatz for $f_{\mu_x}$ and $ f_{\mu_{\varphi}}$ is the $\bar{\mu}$ scheme used in \cite{Chiou:2012pg,Han:2019feb,Han:2020uhb,Han:2022rsx}, which is given by
\begin{eqnarray}\label{BH_mubar}
    f_{\mu_x}(E^x,E^{\varphi}) = \alpha_1 \sqrt{E^x}(E^{\varphi})^{-1}  \, , \qquad f_{\mu_{\varphi}}(E^x,E^{\varphi}) =  \alpha_2 \sqrt{E^x} \,,
\end{eqnarray}
for some constants $ \alpha_1, \alpha_2$ proportional to the minimal area gaps introduced in LQG.

 Note that the conditions (\ref{eq:Choicefmuxfmuphi}-\ref{eq:Choicefmuxfmuphi1}) come from the fact that $K_x$ has a non-trivial density weight, in contrast to $K_{\phi}$ which has density weight $0$. If we exclude the polymerization of $K_x$ and consider the polymerization of $K_{\phi}$ only:
\begin{eqnarray}\label{BH_allowed_eq_2}
   f_1(\mu_x K_x,   \mu_{\varphi} K_{\varphi}) = \tilde{f}_1(\mu_{\varphi} K_{\varphi}) \, , \quad f_2(\mu_x K_x,   \mu_{\varphi} K_{\varphi}) =\mu_x K_x \tilde{f}_2(\mu_{\varphi} K_{\varphi}) \,, %
\end{eqnarray}
e.g. as in models presented in  \cite{Tibrewala:2013kba,Brahma:2014gca,Bojowald:2018xxu,BenAchour:2017ivq,BenAchour:2018khr}.
In this case, we can remove the condition on $f_{\mu_x}$, and a gauge fixing leading to a $\mu_0$-scheme is still allowed.

~\\
The model has five primary constraints
\begin{eqnarray}\label{eq:BH_Pri}
    C_{\text{Pri}} = [ p_{N^x} ,  p_{e_1}, p_{e_2}, p_{\mu_{x}} , p_{\mu_{\varphi}} ]\approx 0.
\end{eqnarray}
Starting from \eqref{PH-sph}, we obtain the following five secondary constraints:
\begin{eqnarray}\label{eq:BH_Sec}
    C_{\text{Sec}} =\{H,   C_{\text{Pri}} \} = [\mathcal{C}^{\rm tot}_x ,  C_{\mu_{x}} , C_{\mu_{\varphi}}, C_{e_1}, C_{e_2} ]\approx 0. 
\end{eqnarray}
Here $C_{\mu_{x},\mu_{\varphi}}$ are the constraints related to implementing the $\mubar$-scheme, they read
\begin{eqnarray}
    C_{\mu_{x}} =\mu_x - f_{\mu_x} \, , \qquad C_{\mu_{\varphi}} = \mu_{\varphi} - f_{\mu_{\varphi}}\,,
\end{eqnarray}
and $C_{e_1,e_2}$ fix $e_1,e_2$ to be:
\begin{eqnarray}
    C_{e_1} &=& e_1 + \frac{4 {E^x} f_2 {\mu_{\varphi}}-K_x {\mu_{x}} \left({E^{\varphi}} {\mu_{x}} \partial_{X}f_1+4 {E^x} {\mu_{\varphi}} \partial_{X}f_2\right)}{2 \sqrt{{E^x}} {\mu_{x}}^2 {\mu_{\varphi}}^2} \, ,\\ \qquad C_{e_2} &=& e_2+\frac{{E^{\varphi}} {\mu_{x}} \left(2 f_1-K_{\varphi} {\mu_{\varphi}} \partial_{Y}f_1\right)+4 {E^x} {\mu_{\varphi}} \left(f_2-K_{\varphi} {\mu_{\varphi}} \partial_{Y}f_2\right)}{2 \sqrt{{E^x}} {\mu_{x}} {\mu_{\varphi}}^3} \,,
\end{eqnarray}
where we defined $X:=\mu_{x} K_{x}, \ Y:= \mu_{\varphi} K_{\varphi}$. 
The stability of $C_{e_1}, C_{e_2},  C_{\mu_{x}} , C_{\mu_{\varphi}}$ can be ensured by choosing  the Lagrangian multipliers $\Lambda_{e_1}, \Lambda_{e_2}, \Lambda_{\mu_x} ,\Lambda_{\mu_{\varphi}}$ appropriately,   while the stability of $\mathcal{C}^{\rm tot}_x$ is automatically satisfied,
\begin{eqnarray}
   \{H,  \mathcal{C}^{\rm tot}_x \} \approx 0 \,.
\end{eqnarray}

The sets $C_{\text{Pri}}$ and $C_{\text{Sec}}$ from \eqref{eq:BH_Pri} and \eqref{eq:BH_Sec} include all constraints of the model, so we end up with ten constraints
\begin{eqnarray*}
C_{\text{all}}= [ p_{N^x} ,  p_{e_1}, p_{e_2}, p_{\mu_{x}} , p_{\mu_{\varphi}},\mathcal{C}^{\rm tot}_x ,  C_{\mu_{x}} , C_{\mu_{\varphi}}, C_{e_1}, C_{e_2} ]\approx 0.   
\end{eqnarray*}
It remains to classify the  constraints (or suitable combinations thereof) as first or second class. We immediately see that $p_{N^x}$ is first class. Since $\mathcal{C}^{\rm tot}_x$ may not be first class as it will have non-trivial brackets with the constraints $C_{\mu_{x}} , C_{\mu_{\varphi}}$ as well as the polymerization functions in $C_{e_1}, C_{e_2},  C_{\mu_{x}} , C_{\mu_{\varphi}}$. However, to circumvent this we introduce an extended diffeomorphism $\tilde{C}^{\rm tot}_x$ on the entire phase space by taking into account the density weight of the individual phase space variables:
\begin{eqnarray}
    \widetilde{\mathcal{C}}^{\rm tot}_x(x)  &=& \mathcal{C}^{\rm tot}_x(x) -2 p_{\mu_x}(x)\mu_x'(x) - p_{\mu_x}'(x)\mu_x(x) - p_{\mu_{\varphi}}(x)\mu_{\varphi}'(x) \nonumber\\
    &&\quad + p_{e_1}(x)e_1'(x) + 2 p_{e_1}'(x)e_1(x) +  p_{e_2}'(x)e_2(x)
\end{eqnarray}
One can verify that $\widetilde{C}^{\rm tot}_x$ is first class, the explicit derivation of this result is presented in appendix \ref{app:diffeo_bracket}. Note that this extension is always possible and does not change the constraint structure because the linear momenta involved in the extra terms are all constraints themselves. The remaining constraints are second class. Hence, from now on we will work with the following set of constraints
\begin{eqnarray*}
\widetilde{C}_{\text{all}}= [ p_{N^x} ,  p_{e_1}, p_{e_2}, p_{\mu_{x}} , p_{\mu_{\varphi}},\widetilde{\mathcal{C}}^{\rm tot}_x ,  C_{\mu_{x}} , C_{\mu_{\varphi}}, C_{e_1}, C_{e_2} ]\approx 0, 
\end{eqnarray*}
including two first class and eight second class constraints.
~\\
In an analogous manner to the cosmological case in section \ref{sec:DiracObsCanTrafo} we will construct a canonical transformation on the extended phase space that allows to perform a Kucha\v{r} decomposition of the extended phase space introduced above. For this purpose we consider the subset of second class constraints $C =\left[  p_{e_1}, p_{e_2}, p_{\mu_{x}} , p_{\mu_{\varphi}}, C_{e_1}, C_{e_2} , C_{\mu_{x}} , C_{\mu_{\varphi}} \right]$.  Similar to the cosmological case, $p_{\mu_{x}} , p_{\mu_{\varphi}}, C_{\mu_{x}} , C_{\mu_{\varphi}}$ do not commute with $C_{e_1}, C_{e_2} $, in addition $C_{e_1}, C_{e_2}$ do not mutually commute. Therefore, as a first step we introduce the following set of variables
\begin{eqnarray*}
\label{eq:CanTrafoSpSymI}
Q^1&:=& C_{e_2},\quad P_1:=p_{e_2},\quad  Q^2:={\cal O}_{C_{e_1}}^{C_{e_2}}={\cal O}^{Q^1}_{C_{e_1}},\quad  P_2:={\cal O}_{C_{p_{e_1}}}^{C_{e_2}}={\cal O}_{C_{p_{e_1}}}^{Q^1}, \\
Q^3&:=& {\cal O}^{C_{e_2},{\cal O}_{C_{e_1}}^{C_{e_2}}}_{C_{\mu_x}}
={\cal O}^{Q^1,Q^2}_{C_{\mu_x}}
,\quad P_3:={\cal O}^{C_{e_2},{\cal O}_{C_{e_1}}^{C_{e_2}}}_{p_{\mu_x}}={\cal O}^{Q^1,Q^2}_{p_{\mu_x}},\\
Q^4&:=&{\cal O}^{C_{e_2},{\cal O}_{C_{e_1}}^{C_{e_2}},{\cal O}^{C_{e_2},{\cal O}_{C_{e_1}}^{C_{e_2}}}_{C_{\mu_x}}}_{C_{\mu_\varphi}}
={\cal O}^{Q^1,Q^2,Q^3}_{C_{\mu_\varphi}},\quad  P_4:={\cal O}^{C_{e_2},{\cal O}_{C_{e_1}}^{C_{e_2}},{\cal O}^{C_{e_2},{\cal O}_{C_{e_1}}^{C_{e_2}}}_{C_{\mu_x}}}_{p_{\mu_\varphi}}={\cal O}^{Q^1,Q^2,Q^3}_{p_{\mu_\varphi}}.
 \end{eqnarray*}
 In the spherically symmetric models the individual dual observable maps can be defined as
 \begin{equation}
 \label{eq:DualObsmapsSphSym}
 {\cal O}^{Q^I}_f(x):=\sum\limits_{n=0}^\infty \int dy_1\dots \int dy_n \int \frac{(P_I(y_1)\cdots P_I(y_n))}{n!}\{\dots \{\{f(x),Q_I(y_1)\},Q_I(y_2)\}\dots \}\},Q_I(y_n)\}. 
\end{equation}
For these subsets of canonical variables we have 
\begin{eqnarray*}
    \{Q^I(x),P_J(y)\} = \delta^I_J \delta(x,y),\quad \{Q^I(x),Q^J(y)\}=0,\quad \{P_I(x),P_J(y)\}=0,\quad I=1\cdots 4.
\end{eqnarray*}
This is explicitly shown in appendic \ref{app:CanTrafo} for the canonical pair $(Q^4,P_4)$. 
If we consider any of the other pairs $(Q^I,P_I)$ the computations work similarly with fewer steps since for the construction of these observables fewer individual observable maps have been applied. As  the computation in appendix \ref{app:CanTrafo} show in these cases the number of observable maps fit again well such that the involved Poisson brackets can be mapped to their corresponding Dirac brackets successively. By construction the  observables $Q^I$ mutually commute. It only remains to show that all observables related to momenta mutually commute and that these commute with all but one $Q^I$. In the two latter cases, the number of applied observable maps for a given $Q^I$ and $P_I$ will not be the same. Therefore, to show that these Poisson brackets vanish works slightly differently in these cases. As an example we consider
\begin{eqnarray*}
\{P_4(x),P_2(y)\} &=& \{ {\cal O}^{Q^1,Q^2,Q^3}_{p_{\mu_\varphi}(x)}, {\cal O}_{{p_{e_1}}}^{Q^1}(y)\} 
=\{ {\cal O}^{Q^2,Q^3,Q^1}_{p_{\mu_\varphi}}(x), {\cal O}_{{p_{e_1}}}^{Q^1}(y)\} 
={\cal O}^{Q^1}_{\{{\cal O}^{Q^2,Q^3}_{p_{\mu_\varphi}}(x),p_{e_1}(y) \}_D^{Q^1}} \\
&=&{\cal O}^{Q^1}_{\{{\cal O}^{Q^2,Q^3}_{p_{\mu_\varphi}}(x),p_{e_1}(y) \}}
\end{eqnarray*}
In the second step we used that the $Q^I$s mutually commute and in the last step that $p_{e_1}$ commutes with $Q^1=C_{e_2},P_1=p_{e_2}$. Finally from 
\begin{equation*}
  \{{\cal O}^{Q^2,Q^3}_{p_{\mu_\varphi}}(x),p_{e_1}(y) \} =\frac{\delta}{\delta e_1(y)} {\cal O}^{Q^2,Q^3}_{p_{\mu_\varphi}}(x) =0 \quad{\rm we\,\, get}\quad \{P_4(x),P_2(y)\}=0. 
\end{equation*}
That the Poisson bracket above vanishes follows from the fact that $e_1(y)$ is involved in $C_{e_1}$ and when it is used in the observable map for other phase space variables than $p_{e_1}$ the part linearly in $e_1(y)$  does not contribute to the nested Poisson bracket. Now we extend the set of canonical pairs in \eqref{eq:CanTrafoSpSymI} by the corresponding observables with respect to $Q^1,\cdots,Q^4$ of the  geometrical and matter degrees of freedom. For this purpose let us introduce the notation ${\cal O}_f^{Q^{1\dots 4}}$ for the observable of the function $f$ with respect to the constraints $Q^{1\cdots 4}$. Then we consider the following canonical transformed variables
\begin{eqnarray*}
\label{eq:CanTrafoSpSymII}
Q^5&:=& {\cal O}_{K_x}^{Q^{1\dots 4}},\quad P_5:={\cal O}_{E^x}^{Q^{1\dots 4}}, \quad Q^6:={\cal O}_{K_\varphi}^{Q^{1\dots 4}},\quad P_6:={\cal O}_{E^\varphi}^{Q^{1\dots 4}},\\
Q^J&:=&{\cal O}_{\phi^A}^{Q^{1\dots 4}},\quad P_J:={\cal O}_{\pi_A}^{Q^{1\dots 4}},\quad J=7,\cdots, L_{\#}+6 
\end{eqnarray*}
As shown in the appendix \ref{app:CanTrafo} with these additional variables we obtain a set of canonical variables satisfying standard CCR given by
\begin{eqnarray*}
 \{Q^I(x),P_J(y)\} =  \widetilde{ \delta}^I_J \delta(x,y),\quad \{Q^I(x),Q^J(y)\} = 0,\quad \{P_I(x),P_J(y)\} =  0,\quad I,J=1,\cdots, L_{\#}+6.   
\end{eqnarray*}
with 
\begin{eqnarray*}
 \widetilde{\delta}^I_J=\delta^I_J \quad {\rm for}\quad I,J=1,\dots, L_{\#}+6 {\quad} {\rm if}\quad I=J\not= 5,6 \quad {\rm and}\quad
 \widetilde{\delta}^5_5=\widetilde{\delta}^6_6=G.
\end{eqnarray*}
In the new coordinates $(Q^I,P_I)$ a Kucha\v{r} decomposition \cite{Kuchar:1972,Hajicek:1999ht} of the kinematical phase space can be achieved. In the explicit computations of the individual Poisson brackets one sees that the reason why they satisfy CCR is very close to the case of the cosmological model just generalized to the field theory case and a larger number of phase space variables. In both cases it turns out to be crucial that the constraints involving the gauge fixing functions $f$ in the cosmological case and $f_{\mu_x},f_{\mu_\varphi}$ are given in a form in which they split into a part depending on the variables $\mu$ and $\mu_x,\mu_\varphi$ respectively and a further part that depends on the gravitational degrees of freedom only, where the latter, in principle, can be generalized to a dependence on the matter degrees of freedom as well. Furthermore, also the fact that $\lambda$ and $e_1,e_2$  respectively only enter linearly in the respective constraints is important.
\subsection{Gauge-fixed model}
\label{sec:GFmodelBH}
Similar to the cosmological model we also discuss the corresponding gauge fixed model in the spherically symmetric case. 
As discussed above the set of constraints $C =\left\{  p_{e_1}, p_{e_2}, p_{\mu_{x}} , p_{\mu_{\varphi}}, C_{e_1}, C_{e_2} , \allowbreak C_{\mu_{x}} , C_{\mu_{\varphi}} \right\}$
forms a second class system. The Dirac matrix has a similar form as \eqref{eq:FullDiracMatrixCosmo}, which reads
\begin{eqnarray}\label{eq:dirac_BH}
   D_{BH} = \left(
\begin{array}{cccccccc}
 0 & 0 & 0 & 0 & -1 & 0 & 0 & 0 \\
 0 & 0 & 0 & 0 & 0 & -1 & 0 & 0 \\
 0 & 0 & 0 & 0 & D_{35} & D_{36} & -1 & 0 \\
 0 & 0 & 0 & 0 & D_{45} & D_{46} & 0 & -1 \\
 1 & 0 & -D_{35} & -D_{45} & 0 & D_{56} & D_{57} & D_{58} \\
 0 & 1 & -D_{36} & -D_{46} & -D_{56} & 0 & D_{67} & D_{68} \\
 0 & 0 & 1 & 0 & -D_{57} & -D_{67} & 0 & D_{78} \\
 0 & 0 & 0 & 1 & -D_{58} & -D_{68} & - D_{78} & 0 \\ 
\end{array}
\right)\, ,
\end{eqnarray}
where the non-trivial components $D_{ij}$ are given explicitly in Appendix \ref{app:BH_Dirac_M}. Note that $D_{78} =0$ if the functions $f_{\mu_x}$ and $f_{\mu_\varphi}$ in \eqref{eq:Choicefmuxfmuphi} do not contain $K_x,K_{\varphi}$.
The inverse of the $D_{BH}$ reads
\begin{eqnarray}
D_{BH}^{-1} = \left(
\begin{array}{cccccccc}
 0 & \tilde{D} & D_{57}-D_{45} D_{78} & D_{58}+D_{35} D_{78} & 1 & 0 & D_{35} & D_{45} \\
 -\tilde{D} & 0 & D_{67}-D_{46} D_{78} & D_{68}+D_{36} D_{78} & 0 & 1 & D_{36} & D_{46} \\
 D_{45} D_{78}-D_{57} & D_{46} D_{78}-D_{67} & 0 & D_{78} & 0 & 0 & 1 & 0 \\
 -D_{58}-D_{35} D_{78} & -D_{68}-D_{36} D_{78} & -D_{78} & 0 & 0 & 0 & 0 & 1 \\
 -1 & 0 & 0 & 0 & 0 & 0 & 0 & 0 \\
 0 & -1 & 0 & 0 & 0 & 0 & 0 & 0 \\
 -D_{35} & -D_{36} & -1 & 0 & 0 & 0 & 0 & 0 \\
 -D_{45} & -D_{46} & 0 & -1 & 0 & 0 & 0 & 0 \\
\end{array}
\right) \nonumber \\
\end{eqnarray}
with 
\begin{eqnarray}
     \tilde{D} = D_{56}+D_{36} D_{57}+D_{46} D_{58}-D_{35} D_{67}-D_{45} D_{68}-D_{36} D_{45} D_{78}+D_{35} D_{46} D_{78}.
\end{eqnarray}
Since $K_x,K_{\varphi}, E^x,E^{\varphi}$ clearly commute with $ p_{e_1}, p_{e_2}, p_{\mu_{x}} , p_{\mu_{\varphi}}$ and we have a vanishing $D_{BH}^{-1}$ for its sub-matrix corresponding to $C_{e_1}, C_{e_2},  C_{\mu_{x}} , C_{\mu_{\varphi}}$, we recover the standard CCR between the gravitational degrees of freedom $K_x,K_{\varphi}, E^x,E^{\varphi}$
\begin{eqnarray}
    &&\{K_x(x),  E^x(y) \}_D = \{K_{\varphi}(x),  E^{\varphi}(y) \}_D = G \delta(x-y) \,,  \\
    &&\{K_x(x),  K_{\varphi}(y) \}_D = \{E^x(x) ,  E^{\varphi}(y) \}_D = 0
\end{eqnarray}
and for matter contributions we have 
\begin{eqnarray}
    &&\{\phi^A(x),\pi_B(y)\}_D=\delta^A_B\delta(x,y) \ , \quad \{\phi^A(x),\phi^B(y)\}_D = \{\pi_A(x),\pi_B(y)\}_D = 0,
\end{eqnarray}
where $A,B\in 1,\dots, L_{\#}$. With the allowed choice of the functions $f_{\mu_x}$ and $f_{\mu_\varphi}$ in \eqref{eq:Choicefmuxfmuphi}, that can be understood as a gauge fixing for $\mu_x$ and $\mu_\varphi$ respectively, the gravitational contribution to the Hamiltonian in the gauge fixed model reads
\begin{eqnarray}
    \mathcal{C}^{\Delta}(x)&=&\frac{1}{2 G}\left[\frac{{{E^{\varphi}}}}{\sqrt{{{E^x}}}}\qty( \frac{f_1\left( \tilde{f}_{\mu_{x}}  \frac{K_{x}}{E^{\varphi}},  \tilde{f}_{\mu_{\varphi}} K_{\varphi}\right)}{\tilde{f}_{\mu_{\varphi}}^2}- 4 E^x\frac{ f_2\left(\tilde{f}_{\mu_{x}}\frac{ K_{x}}{E^{\varphi}},  \tilde{f}_{\mu_{\varphi}}  K_{\varphi}\right)}{\tilde{f}_{\mu_{x}} \tilde{f}_{\mu_{\varphi}}}  + \Gamma^2 - 1  
 )+2\sqrt{E^x} \Gamma'
\right](x) \nonumber
\end{eqnarray}
We can rewrite it as
\begin{eqnarray}\label{eq:gauge_fixed_C_BH}
    \mathcal{C}^{\Delta}(x)&=&\frac{1}{2 G}\left[\frac{{{E^{\varphi}}}}{\sqrt{{{E^x}}}}\qty( f(E^x,K_{\varphi},(E^{\varphi})^{-1} K_x)  + \Gamma^2 - 1  
 )+2\sqrt{E^x} \Gamma'
\right](x) \, ,
\end{eqnarray}
where we combine $f_1,f_2$ and the gauge fixing conditions into a single polymerization function $f(E^x,K_{\varphi},(E^{\varphi})^{-1} K_x) := f_1/\tilde{f}_{\mu_{\varphi}}^2 - 4 E^x f_2/(\tilde{f}_{\mu_{\varphi}}\tilde{f}_{\mu_{x}})$. The polymerization function depends on the combination of $(E^{\varphi})^{-1} K_x$ rather than  depending on either of them only. We remark that, in the gauge fixed model with \eqref{eq:gauge_fixed_C_BH}, we have
\begin{eqnarray}
    \{ \mathcal{C}^{\Delta}(x), \mathcal{C}_x(y)  \}_D = - \partial_x \delta(x -y) \mathcal{C}^{\Delta}(y)
\end{eqnarray}
which confirms that $\mathcal{C}^{\Delta}$ has density weight one. Further requiring scale invariance under the transformation
\begin{eqnarray}
    E^x \to \mu^2 E^x, \quad E^{\varphi} \to \mu^2 E^{\varphi}, \quad K_x \to \mu K_x, \quad K_{\varphi} \to \mu K_{\varphi},
\end{eqnarray}
leads to the following $C^{\Delta}$,
\begin{eqnarray}\label{eq:gauge_fixed_C_BH_new}
    \mathcal{C}^{\Delta}(x)&=&\frac{1}{2 G}\left[\frac{{{E^{\varphi}}}}{\sqrt{{{E^x}}}}\qty( f\left(\frac{K_{\varphi}}{\sqrt{E^x}} ,\frac{\sqrt{E^x} K_x}{E^{\varphi}}  \right)  + \Gamma^2 - 1  
 )+2\sqrt{E^x} \Gamma'
\right](x) \, ,
\end{eqnarray}
which is compatible with the choice of $\mubar$-scheme used in \cite{Chiou:2012pg,Gambini:2020nsf,Kelly:2020uwj,Han:2020uhb,Han:2022rsx}. We note that there are special cases where one or both $f_{\mu_x}$ and $f_{\mu_\varphi}$ are Dirac observables and constants of motion under the effective dynamics, where $f_{\mu_x}$ and $f_{\mu_\varphi}$ can still appear as constants, e.g. similar to the cases in \cite{Ashtekar:2018cay}.
\section{Conclusion}
\label{sec:Concl}
In this article we have shown that we can implement the choice of a  given $\bar{\mu}$-scheme dynamically. Such an implementation corresponds to choosing a phase space dependent function in the argument of a generic polymerization function which replaces holonomies in effective models. We perform the analysis in homogeneous cosmological models as well as spherical symmetric models which are $1+1$-dimensional field theories. To construct the models we extended the usual kinematical phase space by additional canonical pairs which come along with a set of primary constraints so that the number of physical degrees of freedom does not change. These additional constraints build a set of second class constraints for which the corresponding gauge-unfixed model can be constructed in the relational formalism.
The dynamical implementation of a $\bar{\mu}$-scheme can be achieved in terms of a canonical transformation on the extended phase space that allows a Kucha\v{r} decomposition \cite{Kuchar:1972,Hajicek:1999ht} of the extended phase space where all constraints related to the dynamical implementation as well as the Dirac observables with respect to these constraints are the elementary phase space variables.  Furthermore, we showed that in this extended phase space, we can also understand the choice of a given $\mubar$-scheme as a gauge fixing, which gives the polymerized Hamiltonian with the given $\mubar$-scheme in reduced phase space. This gives a possible way from a loop holonomy on a fixed lattice to a polymerization with $\mubar$-scheme. Such extension of the phase space can be included in the path integral formulation of full LQG \cite{Han:2019vpw} to get a $\mubar$-effective dynamics.

In cosmological models, the canonical transformation on the extended phase space and the gauge fixing does not further restrict the choice of gauge fixing function. In particular, the special case of a constant function that corresponds exactly to the $\mu_0$-scheme in LQC is allowed. The $\bar{\mu}$-scheme used in LQC is determined only when we require in addition scale invariance. In contrast, in black hole models, the requirement of keeping the density weight of the scalar Hamiltonian restricts the choice of the gauge fixing functions and hence the possible $\mubar$-schemes. In such case, a constant $\mu_0$-scheme is generally not allowed, unless one does not polymerize the densitized connection $K_x$ carrying a density weight of one, e.g. as in models presented in \cite{Tibrewala:2013kba,Brahma:2014gca,Bojowald:2018xxu,BenAchour:2017ivq,BenAchour:2018khr}. A special case is if the gauge fixing functions themselves corresponding to some constant of motion under the effective dynamics, e.g. similar to the model presented in \cite{Ashtekar:2018lag}. 

In this work we only consider the polymerization of the gravitational variables. The analysis can be generalized to the models with polymerized matter degrees of freedom by introducing additional functions $f_{\mu_{\rm matter}}$. With this, we can go beyond the standard $\mu_0$ and $\mubar$ scheme and allow more general gauge fixing functions that depend on all variables containing both matter and geometry. Our results here, in particular the construction of the canonical transformation on the extended phase space, will work similarly in this more general situation. This becomes relevant for model like for instance in  \cite{Benitez:2020szx,Gambini:2021uzf} in which the matter sector is also polymerized. In these cases, because the momentum of the matter variables is a quantity of density weight one, one could also use it to construct quantities that yield a Hamiltonian with density weight one. Another interesting situation is to extend the analysis to cases beyond spherically symmetric spacetimes. We expect the canonical transformation on the extended phase space and the gauge fixing method will still work, similar to the spherically symmetric case if the involved gauge fixing functions depend on the matter and gravitational degrees of freedom only. In the general case the requirement to have a density weight one Hamiltonian may give us more restrictions on the gauge fixing functions thus constraining the allowed $\mubar$-schemes. In the models studied in this work, the Gau\ss{} constraints have already been gauge fixed. The gauge fixing of the Gau\ss{} constraint is also required in some early attempts to obtain a $\mubar$-scheme effective dynamics from full LQG \cite{Han:2019feb,Han:2021cwb}. It is interesting to generalize the current approach to models that emerge from the LQG theory where the Gau\ss{} constraints have not been gauge fixed. The presence of Gau\ss{} constraints may give strong constraints to the possible form of such gauge fixing conditions. Moreover, in this work we have not considered a polymerization of the diffeomorphism constraints. Such an assumption must be dropped if, for example, we work with models based on a lattice theory on a fixed graph. We expect that work with polymerized diffeomorphism constraints will provide further constraints on admissible $\mubar$-schemes.
\appendix 
\section{$ \tilde{\mathcal{C}}_x^{tot}$ and its Poisson bracket}\label{app:diffeo_bracket}
Here we keep the gauge fixing functions general without imposing \eqref{eq:Choicefmuxfmuphi}, thus the weight of $\mu_x,\mu_{\varphi}$ and $e_1,e_2$ are general. We will derive the correct weight, thus the restriction to gauge fixing functions, such that we have a first class $ \tilde{\mathcal{C}}_x^{tot}$. We start with the following ansatz 
\begin{eqnarray}
    \tilde{\mathcal{C}}_x^{tot}(x)  &=& \mathcal{C}^{\text{tot}}_x(x) + (a-1) p_{\mu_x}(x)\mu_x'(x) +  a p_{\mu_x}'(x)\mu_x(x)+(b-1) p_{\mu_x}(x)\mu_x'(x) + b p_{\mu_x}'(x)\mu_x(x)   \nonumber\\
    &&\quad c p_{e_1}(x)e_1'(x) +(1+c) p_{e_1}'(x)e_1(x) + d p_{e_2}(x)e_2'(x) +(1+d) p_{e_2}'(x)e_2(x)
\end{eqnarray}
The non-trivial Poisson bracket between $\tilde{\mathcal{C}}_x$ and $C_{e_1}, C_{e_2},  C_{\mu_{x}} , C_{\mu_{\varphi}}$ is given by
\begin{eqnarray}
    \{\tilde{C}_x(x), C_{\mu_x}(y) \}&=&-\partial_x  \delta(x-y) \left({a} f_{\mu_{x}}-K_x \partial_{E^x}f_{\mu_{x}}-{E^{\varphi}} \partial_{K_{\phi}}f_{\mu_{x}}\right)(y)\\
    \{\tilde{C}_x(x), C_{\mu_{\varphi}}(y) \}&=&-\partial_x  \delta(x-y) \left({b} f_{\mu_{\varphi}}-K_x \partial_{E^x}f_{\mu_{\varphi}}-{E^{\varphi}} \partial_{K_{\phi}}f_{\mu_{\varphi}}\right)(y)\\
    \{\tilde{C}_x(x), C_{e_1}(y) \}&=&\frac{\partial_x  \delta(x-y)}{{2 \sqrt{{E^x}} f_{\mu_{x}}^2 f_{\mu_{\varphi}}^2}} \Big(4 (1+2 {a}+{b}+{c}) {E^x} f_2 f_{\mu_{\varphi}}-4 {b} {E^x} f_{\mu_{\varphi}}^2 K_{\varphi} \partial_{Y}f_2 \\
    &&-(-1+2 {b}+{c}) {E^{\varphi}} f_{\mu_{x}}^2 K_x \partial_{X}f_1+f_{\mu_{x}} K_x \big(-4 (1+2 {a}+{b}+{c}) {E^x} f_{\mu_{\varphi}} \partial_{X}f_2 \nonumber\\
    &&+{b} {E^{\varphi}} f_{\mu_{x}} f_{\mu_{\varphi}} K_{\varphi} \partial_{X} \partial_{Y}f_1+(1+{a}) {E^{\varphi}} f_{\mu_{x}}^2 K_x \partial_{X}^2f_1 \nonumber\\
    &&+4 {E^x} f_{\mu_{\varphi}} \left({b} f_{\mu_{\varphi}} K_{\varphi} \partial_{X} \partial_{Y}f_2+(1+{a}) f_{\mu_{x}} K_x \partial_{X}^2f_2\right)\big)\Big)(y) \nonumber\\
   \{\tilde{C}_x(x), C_{e_2}(y) \}&=&\frac{\partial_x  \delta(x-y) }{2 \sqrt{{E^x}} f_{\mu_{x}} f_{\mu_{\varphi}}^3}\Big({E^{\varphi}} f_{\mu_{x}} \big(2 (3 {b}+{d}) f_1-2 (1+{a}) f_{\mu_{x}} K_x \partial_{X}f_1\\
&&+f_{\mu_{\varphi}} K_{\varphi} \left(-\left((4 {b}+{d}) \partial_{Y}f_1\right)+{b} f_{\mu_{\varphi}} K_{\varphi} \partial_{Y}^2f_1+(1+{a}) f_{\mu_{x}} K_x \partial_{X} \partial_{Y}f_1\right)\big) \nonumber\\
&&+4 {E^x} f_{\mu_{\varphi}} \big((1+{a}+2 {b}+{d}) f_2-(1+{a}+2 {b}+{d}) f_{\mu_{\varphi}} K_{\varphi} \partial_{Y}f_2 \nonumber\\
&&+{b} f_{\mu_{\varphi}}^2 K_{\varphi}^2 \partial_{Y}^2f_2+(1+{a}) f_{\mu_{x}} K_x \left(-\partial_{X}f_2+f_{\mu_{\varphi}} K_{\varphi} \partial_{X} \partial_{Y}f_2\right)\big)\Big)(y) \nonumber
\end{eqnarray}
We notice that, in order to have $\{\tilde{\mathcal{C}}_x, C_{e_1, e_2}\} \approx 0 $ for generic $f_1,f_2$, we need 
\begin{eqnarray}
    a = -1\ , \quad  b= 0 \ , \quad c = 1 \ , \quad d = 0
\end{eqnarray}
With this set of values, $\mu_x$ has density weight $-1$ and $\mu_{\varphi}$ has density weight $0$. $\{\tilde{\mathcal{C}}_x, C_{\mu_x, \mu_{\varphi}}\}$ now becomes:
\begin{eqnarray}
    \{\tilde{\mathcal{C}}_x(x), C_{\mu_x, \mu_{\varphi}}(y)\} &=& \partial_x \delta(x-y) \left( f_{\mu_x} + K_x \partial_{E^x}f_{\mu_{x}}-{E^{\varphi}} \partial_{K_{\phi}}f_{\mu_{x}}  \right)  (y) \ ,\\
    \{\tilde{\mathcal{C}}_x(x), C_{\mu_x, \mu_{\varphi}}(y)\} &=& \partial_x \delta(x-y) \left( K_x \partial_{E^x}f_{\mu_{\varphi}}-{E^{\varphi}} \partial_{K_{\phi}}f_{\mu_{\varphi}} \right)  (y)
\end{eqnarray}
Requiring the right hand side to vanish gives the condition \eqref{eq:Choicefmuxfmuphi}, which comes from requiring $C^{\Delta}$ have the correct density weight one. With this condition one can check easily
\begin{eqnarray}
    \{\tilde{C}_x(x), C^{\Delta}_{tot} (y)\} = \partial_x \delta(x-y) C^{\Delta}(y)
\end{eqnarray}
where we use $\{\tilde{C}_x^{\text{matter}}(x), C^{\text{matter}} (y)\} = \partial_x \delta(x-y) C^{\text{matter}}(y)$.
As a result, we also have $\{H,\tilde{\mathcal{C}}_x^{tot}\} = 0$. With this fact we finish the constraint analysis with
$p_{N^x}, \tilde{\mathcal{C}}_x^{tot}$ are first class, and  $C =\left\{  p_{e_1}, p_{e_2}, p_{\mu_{x}} , p_{\mu_{\varphi}}, C_{\mu_{x}} , C_{\mu_{\varphi}}, C_{e_1}, C_{e_2}  \right\}$
form a second class system.

\section{Explicit form of Dirac Matrix in BH}\label{app:BH_Dirac_M}
Here we give in detail the elements $D_{ij}=D_{ij}\delta(x-y)$ of Dirac matrix presented in \eqref{eq:dirac_BH}. Since there is no partial derivatives appearing in the second class constraint, the following elements are the same as these of the Dirac matrix for homogeneous Kontowski-Sachs spacetime.
\begin{eqnarray*}
    D_{35}&=& \frac{8 {E^x} f_2 {\mu_{\varphi}}+K_x {\mu_{x}} \left(-8 {E^x} {\mu_{\varphi}} \partial_{X}f_2+K_x {\mu_{x}} \left({E^{\varphi}} {\mu_{x}} \partial_{X}^2f_1+4 {E^x} {\mu_{\varphi}} \partial_{X}^2f_2\right)\right)}{2 \sqrt{{E^x}} {\mu_{x}}^3 {\mu_{\varphi}}^2} \\
    D_{36}&=& \frac{{E^{\varphi}} K_x \left(-2 \partial_{X}f_1+K_{\varphi} {\mu_{\varphi}} \partial_{X} \partial_{Y}f_1\right)}{2 \sqrt{{E^x}} {\mu_{\varphi}}^3} \nonumber \\
    &&+\frac{2 \sqrt{{E^x}}\left(f_2-K_{\varphi} {\mu_{\varphi}} \partial_{Y}f_2-K_x {\mu_{x}} \partial_{X}f_2+K_x K_{\varphi} {\mu_{x}} {\mu_{\varphi}} \partial_{X} \partial_{Y}f_2\right)}{ {\mu_{x}}^2 {\mu_{\varphi}}^2}\\
    D_{45}&=& \frac{{E^{\varphi}} K_x \left(-2 \partial_{X}f_1+K_{\varphi} {\mu_{\varphi}} \partial_{X} \partial_{Y}f_1\right)}{2 \sqrt{{E^x}}  {\mu_{\varphi}}^3} \\
    && + \frac{2 \sqrt{{E^x}} {\mu_{\varphi}} \left(f_2-K_{\varphi} {\mu_{\varphi}} \partial_{Y}f_2-K_x {\mu_{x}} \partial_{X}f_2+K_x K_{\varphi} {\mu_{x}} {\mu_{\varphi}} \partial_{X} \partial_{Y}f_2\right)}{2 \sqrt{{E^x}} {\mu_{x}}^2 {\mu_{\varphi}}^2} \\
    D_{46}&=& \frac{{E^{\varphi}} \left(6 f_1+K_{\varphi} {\mu_{\varphi}} \left(-4 \partial_{Y}f_1+K_{\varphi} {\mu_{\varphi}} \partial_{Y}^2f_1\right)\right)}{2 \sqrt{{E^x}} {\mu_{\varphi}}^4} \\
    && + \frac{2 \sqrt{{E^x}} \left(2 f_2+K_{\varphi} {\mu_{\varphi}} \left(-2 \partial_{Y}f_2+K_{\varphi} {\mu_{\varphi}} \partial_{Y}^2f_2\right)\right)}{{\mu_{x}} {\mu_{\varphi}}^3} \\
    D_{56}&=& \frac{-2 {E^x} K_x {\mu_{x}} {\mu_{\varphi}} \left(-{E^{\varphi}} {\mu_{x}} \partial_{Y}f_1+K_{\varphi} {\mu_{\varphi}} \left({E^{\varphi}} {\mu_{x}} \partial_{Y}^2f_1+4 {E^x} {\mu_{\varphi}} \partial_{Y}^2f_2\right)\right) \partial_{X}f_1}{8 {E^x}^2 {\mu_{x}}^2 {\mu_{\varphi}}^5} \\
    &&+\frac{2 {E^x} {\mu_{\varphi}} \left(-2 f_1+K_{\varphi} {\mu_{\varphi}} \partial_{Y}f_1\right) \left(-4 {E^x} {\mu_{\varphi}} \partial_{Y}f_2+K_x {\mu_{x}} \left({E^{\varphi}} {\mu_{x}} \partial_{X} \partial_{Y}f_1+4 {E^x} {\mu_{\varphi}} \partial_{X} \partial_{Y}f_2\right)\right)}{8 {E^x}^2 {\mu_{x}}^2 {\mu_{\varphi}}^5} \\
    &&+\frac{\left({E^{\varphi}} K_x {\mu_{x}}^2 \partial_{X}f_1+4 {E^x} {\mu_{\varphi}} \left(f_2-K_x {\mu_{x}} \partial_{X}f_2\right)\right)}{8 {E^x}^2 {\mu_{x}}^2 {\mu_{\varphi}}^5} \times \\
     && \left(-2 {E^{\varphi}} {\mu_{x}} \partial_{X}f_1+{\mu_{\varphi}} \left(-4 {E^x} \partial_{X}f_2+{E^{\varphi}} K_{\varphi} {\mu_{x}} \partial_{X} \partial_{Y}f_1+4 {E^x} K_{\varphi} {\mu_{\varphi}} \partial_{X} \partial_{Y}f_2\right)\right) \\
    &&+\frac{\left({E^{\varphi}} \partial_{X}f_1+{E^{\varphi}} K_x {\mu_{x}} \partial_{X}^2f_1+4 {E^x} K_x {\mu_{\varphi}} \partial_{X}^2f_2\right)}{8 {E^x}^2 {\mu_{x}} {\mu_{\varphi}}^5} \times \\
    &&\left({E^{\varphi}} {\mu_{x}} \left(2 f_1-K_{\varphi} {\mu_{\varphi}} \partial_{Y}f_1\right)+4 {E^x} {\mu_{\varphi}} \left(-f_2+K_{\varphi} {\mu_{\varphi}} \partial_{Y}f_2\right)\right) \\
    D_{57}&=&\frac{-4 {E^x} {\mu_{\varphi}} \partial_{Y}f_2+K_x {\mu_{x}} \left({E^{\varphi}} {\mu_{x}} \partial_{X} \partial_{Y}f_1+4 {E^x} {\mu_{\varphi}} \partial_{X} \partial_{Y}f_2\right) }{2 \sqrt{{E^x}} {\mu_{x}}^2 {\mu_{\varphi}}}\partial_{E^{\varphi}}{\tilde{f}_{\mu_x}}\\
    &&+ \frac{{E^{\varphi}} K_x {\mu_{x}}^2 \partial_{X}f_1+4 {E^x} {\mu_{\varphi}} \left(f_2-K_x {\mu_{x}} \partial_{X}f_2\right) }{4 {E^x}^{3/2} {\mu_{x}}^2 {\mu_{\varphi}}^2}\partial_{K_{x}}{\tilde{f}_{\mu_x}}\\
    &&+ \frac{{E^{\varphi}} \partial_{X}f_1+{E^{\varphi}} K_x {\mu_{x}} \partial_{X}^2f_1+4 {E^x} K_x {\mu_{\varphi}} \partial_{X}^2f_2 }{2 \sqrt{{E^x}} {\mu_{\varphi}}^2} \partial_{E^x}{\tilde{f}_{\mu_x}} -\frac{K_x \partial_{X}f_1 \partial_{K_{\varphi}}{\tilde{f}_{\mu_x}}}{2 \sqrt{{E^x}} {\mu_{\varphi}}^2}\\
     D_{58}&=&\frac{-4 {E^x} {\mu_{\varphi}} \partial_{Y}f_2+K_x {\mu_{x}} \left({E^{\varphi}} {\mu_{x}} \partial_{X} \partial_{Y}f_1+4 {E^x} {\mu_{\varphi}} \partial_{X} \partial_{Y}f_2\right)}{2 \sqrt{{E^x}} {\mu_{x}}^2 {\mu_{\varphi}}} \partial_{E^{\varphi}}\tilde{f}_{\mu_{\varphi}}\\
     &&+\frac{{E^{\varphi}} K_x {\mu_{x}}^2 \partial_{X}f_1+4 {E^x} {\mu_{\varphi}} \left(f_2-K_x {\mu_{x}} \partial_{X}f_2\right)}{4 {E^x}^{3/2} {\mu_{x}}^2 {\mu_{\varphi}}^2} \partial_{K_{x}}\tilde{f}_{\mu_{\varphi}}\\
     &&+\frac{{E^{\varphi}} \partial_{X}f_1+{E^{\varphi}} K_x {\mu_{x}} \partial_{X}^2f_1+4 {E^x} K_x {\mu_{\varphi}} \partial_{X}^2f_2 }{2 \sqrt{{E^x}} {\mu_{\varphi}}^2} \partial_{E^x}\tilde{f}_{\mu_{\varphi}} -\frac{K_x \partial_{X}f_1 \partial_{K_{\varphi}}\tilde{f}_{\mu_{\varphi}}}{2 \sqrt{{E^x}} {\mu_{\varphi}}^2} \\
     D_{67}&=&\frac{-{E^{\varphi}} {\mu_{x}} \partial_{Y}f_1+K_{\varphi} {\mu_{\varphi}} \left({E^{\varphi}} {\mu_{x}} \partial_{Y}^2f_1+4 {E^x} {\mu_{\varphi}} \partial_{Y}^2f_2\right) }{2 \sqrt{{E^x}} {\mu_{x}} {\mu_{\varphi}}^2} \partial_{E^{\varphi}}{\tilde{f}_{\mu_x}}+\frac{2 f_1-K_{\varphi} {\mu_{\varphi}} \partial_{Y}f_1 }{2 \sqrt{{E^x}} {\mu_{\varphi}}^3}\partial_{K_{\varphi}}{\tilde{f}_{\mu_x}}\\
     &&\frac{-2 {E^{\varphi}} f_1 {\mu_{x}}+4 {E^x} f_2 {\mu_{\varphi}}+K_{\varphi} {\mu_{\varphi}} \left({E^{\varphi}} {\mu_{x}} \partial_{Y}f_1-4 {E^x} {\mu_{\varphi}} \partial_{Y}f_2\right) }{4 {E^x}^{3/2} {\mu_{x}} {\mu_{\varphi}}^3} \partial_{K_{x}}{\tilde{f}_{\mu_x}}\\
     &&\frac{-2 {E^{\varphi}} {\mu_{x}} \partial_{X}f_1+{\mu_{\varphi}} \left(-4 {E^x} \partial_{X}f_2+{E^{\varphi}} K_{\varphi} {\mu_{x}} \partial_{X} \partial_{Y}f_1+4 {E^x} K_{\varphi} {\mu_{\varphi}} \partial_{X} \partial_{Y}f_2\right) }{2 \sqrt{{E^x}} {\mu_{\varphi}}^3}\partial_{E^x}{\tilde{f}_{\mu_x}}
     \\
     D_{68}&=&\frac{-{E^{\varphi}} {\mu_{x}} \partial_{Y}f_1+K_{\varphi} {\mu_{\varphi}} \left({E^{\varphi}} {\mu_{x}} \partial_{Y}^2f_1+4 {E^x} {\mu_{\varphi}} \partial_{Y}^2f_2\right) }{2 \sqrt{{E^x}} {\mu_{x}} {\mu_{\varphi}}^2}\partial_{E^{\varphi}}\tilde{f}_{\mu_{\varphi}}+\frac{2 f_1-K_{\varphi} {\mu_{\varphi}} \partial_{Y}f_1}{2 \sqrt{{E^x}} {\mu_{\varphi}}^3}\partial_{K_{\varphi}}\tilde{f}_{\mu_{\varphi}}\\
     &&\frac{-2 {E^{\varphi}} f_1 {\mu_{x}}+4 {E^x} f_2 {\mu_{\varphi}}+K_{\varphi} {\mu_{\varphi}} \left({E^{\varphi}} {\mu_{x}} \partial_{Y}f_1-4 {E^x} {\mu_{\varphi}} \partial_{Y}f_2\right)}{4 {E^x}^{3/2} {\mu_{x}} {\mu_{\varphi}}^3}\partial_{K_{x}}\tilde{f}_{\mu_{\varphi}}\\
     &&\frac{-2 {E^{\varphi}} {\mu_{x}} \partial_{X}f_1+{\mu_{\varphi}} \left(-4 {E^x} \partial_{X}f_2+{E^{\varphi}} K_{\varphi} {\mu_{x}} \partial_{X} \partial_{Y}f_1+4 {E^x} K_{\varphi} {\mu_{\varphi}} \partial_{X} \partial_{Y}f_2\right)}{2 \sqrt{{E^x}} {\mu_{\varphi}}^3}\partial_{E^x}\tilde{f}_{\mu_{\varphi}}\\
     D_{78}&=&-\partial_{E^{\phi}}\tilde{f}_{\mu_{\varphi}} \partial_{K_{\phi}}\tilde{f}_{\mu_{x}}+\partial_{E^{\phi}}\tilde{f}_{\mu_{x}} \partial_{K_{\phi}}\tilde{f}_{\mu_{\varphi}}-\partial_{E^x}\tilde{f}_{\mu_{\varphi}} \partial_{K_x}\tilde{f}_{\mu_{x}}+\partial_{E^x}\tilde{f}_{\mu_{x}} \partial_{K_x}\tilde{f}_{\mu_{\varphi}}%
\end{eqnarray*}

\section{Proof of Lemma 1}
\label{app:Lemma}
In this appendix we present the proof of lemma 1 used in the main text. For the benefit of the reader we also show lemma one here again
\begin{customlemma}{1}
\label{lemma1}
For the iterated Poisson bracket of observables we have
\begin{equation}
\{ {\cal O}^{C_\lambda}_f, {\cal O}^{C_\lambda}_g \}_{(n)} = {\cal O}^{C_\lambda}_{\{f,g\}_{D(n)}^{C_\lambda}}.
\end{equation}
\end{customlemma}
\begin{customproof}{1}
We start with the case  n=2 for which we have
\begin{equation*}
\{ {\cal O}^{C_\lambda}_f, {\cal O}^{C_\lambda}_g \}_{(2)}
=\{\{ {\cal O}^{C_\lambda}_f, {\cal O}^{C_\lambda}_g \}, {\cal O}^{C_\lambda}_g\} 
=\{ {\cal O}^{C_\lambda}_{\{f,g\}_D^{C_\lambda}},  {\cal O}^{C_\lambda}_g\} =  {\cal O}^{C_\lambda}_{\{\{f,g\}_D^{C_\lambda},g\}_D^{C_\lambda}
}={\cal O}^{C_\lambda}_{\{f,g\}_{D(2)}^{C_\lambda}}
\end{equation*}
In the second step we used 
\begin{equation*}
\{ {\cal O}^{C_\lambda}_f, {\cal O}^{C_\lambda}_g \}_{(1)}
=\{ {\cal O}^{C_\lambda}_f, {\cal O}^{C_\lambda}_g \} = {\cal O}^{C_\lambda}_{\{f,g\}^{C_\lambda}_D}
\end{equation*}
which has been proven in \cite{Thiemann:2004wk} involving a weak equality sign there. In our case  due to $\{C_\lambda,p_\lambda\}=1$ this becomes a strong equality here. Now we assume that the statement is true for generic $n$ and perform the induction step. We obtain
\begin{eqnarray*}
\{ {\cal O}^{C_\lambda}_f, {\cal O}^{C_\lambda}_g \}_{(n+1)}&=&
\{\{{\cal O}^{C_\lambda}_f, {\cal O}^{C_\lambda}_g\}_{(n)}, {\cal O}^{C_\lambda}_g\} 
=\{ {\cal O}^{C_\lambda}_{\{f,g\}_{D(n)}^{C_\lambda}}, {\cal O}^{C_\lambda}_g\}  \\
&=&
{\cal O}^{C_\lambda}_{\{\{f,g\}_{D(n)}^{C_\lambda}, g\}_D^{C_\lambda}} = {\cal O}^{C_\lambda}_{\{f,g\}_{D(n+1)}^{C_\lambda}} \qed
\end{eqnarray*}
\end{customproof}

\section{Kucha\v{r} decomposition for the spherically symmetric model}
\label{app:CanTrafo}
In this appendix we present more details how the Kuchar decomposition can be explicitly constructed. For this purpose first we show that $\{ Q^I(x), P_J(y) \} =\delta(x,y)$ for $I,J=1,\dots,4$ and afterwards we will we show that $\{Q^I(x),P_J(y)\}$ for $I,J=5\dots N+6$ satisfy standard CCR and that $Q^I,P_I$ for $I=5\dots N+6$ commute with all canonical pairs $(Q^I,P_I)$ for $I=1\dots 4$. In both cases we will show this for one case only the remaining cases work exactly similar. For the first case we consider $Q^4,P_4$ and we have
\begin{eqnarray*}
 \{ Q^4(x), P_4(y) \} &=& \{ {\cal O}^{Q^1,Q^2,Q^3}_{C_{\mu_\varphi}}(x), {\cal O}^{Q^1,Q^2,Q^3}_{p_{\mu_\varphi}}(y)\}
 =  {\cal O}^{Q^3}_{\{ {\cal O}^{Q^1,Q^2}_{C_{\mu_\varphi}}(x)\, , \,{\cal O}^{Q^1,Q^2}_{p_{\mu_\varphi}}(y)\}_{D}^{Q^3}}. 
\end{eqnarray*}
The involved Dirac bracket is given by
\begin{eqnarray}
\label{eq:DBQ3}
 \{ {\cal O}^{Q^1,Q^2}_{C_{\mu_\varphi}}(x)\, , \,{\cal O}^{Q^1,Q^2}_{p_{\mu_\varphi}}(y)\}_{D}^{Q^3} 
 &=&
 \{ {\cal O}^{Q^1,Q^2}_{C_{\mu_\varphi}}(x)\, , \,{\cal O}^{Q^1,Q^2}_{p_{\mu_\varphi}}(y)\} -
\int dz \{ {\cal O}^{Q^1,Q^2}_{C_{\mu_\varphi}}(x)\, ,\, Q^3(z)\}\{P_3(z)\, , \, {\cal O}^{Q^1,Q^2}_{p_{\mu_\varphi}}(y) \} \nonumber\\
&& +\int dz \{ {\cal O}^{Q^1,Q^2}_{C_{\mu_\varphi}}(x)\, ,\, P_3(z)\}\{Q^3(z)\, , \, {\cal O}^{Q^1,Q^2}_{p_{\mu_\varphi}}(y) \}.
\end{eqnarray}
We further have
\begin{eqnarray*}
 \{ {\cal O}^{Q^1,Q^2}_{C_{\mu_\varphi}}(x)\, ,\, P_3(z)\}
 &=& 
 {\cal O}^{Q^2}_{\{{\cal O}^{Q^1}_{C_{\mu_\varphi}}(x)
\, , \, {\cal O}^{Q^1}_{p_{\mu_x}}(z)\}_D^{Q^2}} \\
\end{eqnarray*}
 with
 \begin{eqnarray}
  \label{eq:DBQ2}
\{ {\cal O}^{Q^1}_{C_{\mu_\varphi}}(x)
\, , \, {\cal O}^{Q^1}_{p_{\mu_x}}(z)\}_D^{Q^2} &=&
 \{ {\cal O}^{Q^1}_{C_{\mu_\varphi}}(x)\, , \,{\cal O}^{Q^1}_{p_{\mu_x}}(y)\} -
\int dz \{ {\cal O}^{Q^1}_{C_{\mu_\varphi}}(x)\, ,\, Q^2(z)\}\{P_2(z)\, , \, {\cal O}^{Q^1}_{p_{\mu_\varphi}}(y) \} \nonumber\\
&& +\int dz \{ {\cal O}^{Q^1}_{C_{\mu_\varphi}}(x)\, ,\, P_2(z)\}\{Q^2(z)\, , \, {\cal O}^{Q^1}_{p_{\mu_\varphi}}(y) \}\nonumber .
 \end{eqnarray}
 Now we can use 
\begin{eqnarray*}
 \{ {\cal O}^{Q^1}_{C_{\mu_\varphi}}(x)\, ,\, P_2(z)\}
 &=& {\cal O}^{Q^1}_{
 \{{C_{\mu_\varphi}}(x), p_{e_1}(z) \}^{Q^1}_{D}}
 \end{eqnarray*}
 with
 \begin{eqnarray*}
\{{C_{\mu_\varphi}}(x), p_{e_1}(z) \}^{Q^1}_{D}     
&=& \{{C_{\mu_\varphi}}(x), p_{e_1}(z) \}
 -
\int dz' \{ C_{\mu_\varphi}(x), C_{e_2}(z') \}\{p_{e_2}(z'), p_{e_1}(z) \} \\
&&+ \int dz' \{ C_{\mu_\varphi}(x), p_{e_2}(z') \}\{ C_{e_2}(z'), p_{e_1}(z) \} \\
&=&0.
\end{eqnarray*}
Likewise we obtain $\{P_2(z)\, , \, {\cal O}^{Q^1}_{p_{\mu_\varphi}}(y)\}=\{ {\cal O}^{Q^1}_{C_{\mu_\varphi}}(x)\, , \,{\cal O}^{Q^1}_{p_{\mu_x}}(y)\}=0$. Hence, we obtain 
\begin{equation*}
\{ {\cal O}^{Q^1,Q^2}_{C_{\mu_\varphi}}(x)\, ,\, P_3(z)\} =0 \ ,\quad \{P_3(z)\, , \, {\cal O}^{Q^1,Q^2}_{p_{\mu_\varphi}}(y) \}=0,
\end{equation*}
Using this in \eqref{eq:DBQ3} we have
\begin{equation*}
\{ {\cal O}^{Q^1,Q^2}_{C_{\mu_\varphi}}(x)\, , \,{\cal O}^{Q^1,Q^2}_{p_{\mu_\varphi}}(y)\}_{D}^{Q^3} 
= \{ {\cal O}^{Q^1,Q^2}_{C_{\mu_\varphi}}(x)\, , \,{\cal O}^{Q^1,Q^2}_{p_{\mu_\varphi}}(y)\} 
\end{equation*}
Finally, we consider the remaining Poisson bracket 
\begin{eqnarray*}
\{ {\cal O}^{Q^1,Q^2}_{C_{\mu_\varphi}}(x)\, , \,{\cal O}^{Q^1,Q^2}_{p_{\mu_\varphi}}(y)\} 
&=&{\cal O}^{Q^2}_{ \{ {\cal O}^{Q^1}_{C_{\mu_\varphi}}(x)\, , \,{\cal O}^{Q^1}_{p_{\mu_\varphi}}(y)\}^{Q^2}_D }
\end{eqnarray*}
with
\begin{eqnarray*}
\{ {\cal O}^{Q^1}_{C_{\mu_\varphi}}(x)\, , \,{\cal O}^{Q^1}_{p_{\mu_\varphi}}(y)\}^{Q^2}_D &=& \{ C_{\mu_\varphi}(x)\, , \, p_{\mu_\varphi}(y)\}^{Q^1}_D
 -
\int dz \{ C_{\mu_\varphi}(x), C_{e_2}(z) \}^{Q^1}_D\{p_{e_2}(z), p_{\mu_\varphi}(y) \}^{Q^1}_{D} \\
&&+ \int dz \{ C_{\mu_\varphi}(x), p_{e_2}(z) \}^{Q^1}_D\{ e_2(z), p_{\mu_\varphi}(y) \}^{Q^1}_D \\
&=&
\{ C_{\mu_\varphi}(x)\, , \, p_{\mu_\varphi}(y)\}=\delta(x,y).
\end{eqnarray*}
Given this and reinserting it back into \eqref{eq:DBQ3} we finally end up with
\begin{eqnarray*}
 \{Q^4(x),P_4(y)\} &=& {\cal O}^{Q^3}_{\delta(x,y)} = \delta(x,y).  
\end{eqnarray*}

Next we want to show that $\{Q^I(x),P_J(y)\}$ for $I,J=5\dots N+6$ satisfy standard CCR and that $Q^I,P_I$ for $I=5\dots N+6$ commute with all canonical pairs $(Q^I,P_I)$ for $I=1\dots 4$. Here we consider the canonical pair $(Q^5,P_5)$ as an example that we discuss in detail. The remaining combinations of variables work similarly. We start with
\begin{eqnarray*}
 \{Q^5(x), P_5(y)\} &=&    \{{\cal O}^{Q^{1\dots 4}}_{K_x}(x), {\cal O}^{Q^{1\dots 4}}_{E^x}(y) \}
 ={\cal O}^{Q^4}_{\{{\cal O}^{Q^{1\dots 3}}_{K_x}(x), {\cal O}^{Q^{1\dots 3}}_{E^x}(y) \}_D^{Q^4}}
\end{eqnarray*}
We have using $Q^4={\cal O}^{Q^{1\dots 3}}_{C_{\mu_\varphi}}$ and $P_4={\cal O}^{Q^{1\dots 3}}_{p_{\mu_\varphi}}$
\begin{eqnarray*}
\{{\cal O}^{Q^{1\dots 3}}_{K_x}(x), {\cal O}^{Q^{1\dots 3}}_{E^x}(y) \}_D^{Q^4}&=&
\{{\cal O}^{Q^{1\dots 3}}_{K_x}(x), {\cal O}^{Q^{1\dots 3}}_{E^x}(y) \}
-\int dz \{{\cal O}^{Q^{1\dots 3}}_{K_x}(x) , {\cal O}^{Q^{1\dots 3}}_{C_{\mu_\varphi}}(z)\}
\{ {\cal O}^{Q^{1\dots 3}}_{p_{\mu_\varphi}}(z) ,{\cal O}^{Q^{1\dots 3}}_{E^x}(y) \} \\
&&
+\int dz \{{\cal O}^{Q^{1\dots 3}}_{K_x}(x) , {\cal O}^{Q^{1\dots 3}}_{p_{\mu_\varphi}}(z)\}
\{ {\cal O}^{Q^{1\dots 3}}_{C_{\mu_\varphi}}(z) ,{\cal O}^{Q^{1\dots 3}}_{E^x}(y) \}.
\end{eqnarray*}
Next we want to show that the Poisson brackets involving ${\cal O}^{Q^{1\dots 3}}_{p_{\mu_\varphi}}$ all vanish. For this purpose we consider the following as an example 
\begin{eqnarray*}
 \{ {\cal O}^{Q^{1\dots 3}}_{p_{\mu_\varphi}}(z) ,{\cal O}^{Q^{1\dots 3}}_{E^x}(y) \}
 &=&
 {\cal O}^{Q^3}_{\{{\cal O}^{Q^{1\dots 2}}_{p_{\mu_\varphi}}(z), {\cal O}^{Q^{1\dots 2}}_{E^x}(y) \}_D^{Q^3}} 
 \end{eqnarray*}
 with
 \begin{eqnarray*}
 \{{\cal O}^{Q^{1\dots 2}}_{p_{\mu_\varphi}}(z), {\cal O}^{Q^{1\dots 2}}_{E^x}(y) \}_D^{Q^3}    
 &=&
 \{{\cal O}^{Q^{1\dots 2}}_{p_{\mu_\varphi}}(z), {\cal O}^{Q^{1\dots 2}}_{E^x}(y) \}
-\int dz' \{{\cal O}^{Q^{1\dots 2}}_{p_{\mu_\varphi}}(z) , {\cal O}^{Q^{1\dots 2}}_{C_{\mu_x}}(z')\}
\{ {\cal O}^{Q^{1\dots 2}}_{p_{\mu_x}}(z') ,{\cal O}^{Q^{1\dots 2}}_{E^x}(y) \} \\
&&
+\int dz' \{{\cal O}^{Q^{1\dots 2}}_{p_{\mu_\varphi}}(z), {\cal O}^{Q^{1\dots 2}}_{p_{\mu_x}}(z')\}
\{ {\cal O}^{Q^{1\dots 2}}_{C_{\mu_x}}(z') ,{\cal O}^{Q^{1\dots 2}}_{E^x}(y) \},
\end{eqnarray*}
where we used $Q^3={\cal O}^{Q^{1\dots 2}}_{C_{\mu_x}}$ and $P_3={\cal O}^{Q^{1\dots 2}}_{p_{\mu_x}}$.
Here we consider the Poisson brackets involving ${\cal O}^{Q^{1\dots 2}}_{p_{\mu_x}}(z')$ further and get
\begin{eqnarray*}
 \{ {\cal O}^{Q^{1\dots 2}}_{p_{\mu_x}}(z') ,{\cal O}^{Q^{1\dots 2}}_{E^x}(y) \}
 &=&
 {\cal O}^{Q^2}_{\{{\cal O}^{Q^{1}}_{p_{\mu_x}}(z'), {\cal O}^{Q^{1}}_{E^x}(y) \}_D^{Q^2}} 
 \end{eqnarray*}
 with
 \begin{eqnarray*}
 \{{\cal O}^{Q^{1}}_{p_{\mu_x}}(z'), {\cal O}^{Q^{1}}_{E^x}(y) \}_D^{Q^2}    
 &=&
 \{{\cal O}^{Q^{1}}_{p_{\mu_x}}(z'), {\cal O}^{Q^{1}}_{E^x}(y) \}
-\int dz^{\prime\prime} \{{\cal O}^{Q^{1}}_{p_{\mu_x}}(z') , {\cal O}^{Q^{1}}_{C_{e_1}}(z^{\prime\prime})\}
\{ {\cal O}^{Q^{1}}_{p_{e_1}}(z^{\prime\prime}) ,{\cal O}^{Q^{1}}_{E^x}(y) \} \\
&&
+\int dz^{\prime\prime} \{{\cal O}^{Q^{1}}_{p_{\mu_x}}(z^{\prime}), {\cal O}^{Q^{1}}_{p_{e_1}}(z^{\prime\prime})\}
\{ {\cal O}^{Q^{1}}_{C_{e_1}}(z^{\prime\prime}) ,{\cal O}^{Q^{1}}_{E^x}(y) \}.
\end{eqnarray*}
Given the last equation it is easy to show that $ \{{\cal O}^{Q^{1}}_{p_{\mu_x}}(z'), {\cal O}^{Q^{1}}_{E^x}(y) \}_D^{Q^2} =0$ because we have 
\begin{eqnarray*}
  \{{\cal O}^{Q^{1}}_{p_{\mu_x}}(z'), {\cal O}^{Q^{1}}_{E^x}(y) \} &=&
  {\cal O}^{Q^{1}}_{ \{ p_{\mu_x}(z'),E^x(y) \}^{Q^1}_D} =0 \\
  \{ {\cal O}^{Q^{1}}_{p_{e_1}}(z^{\prime\prime}) ,{\cal O}^{Q^{1}}_{E^x}(y) \}
  &=&{\cal O}^{Q^{1}}_{ \{ p_{e_1}(z^{\prime\prime}),E^x(y) \}^{Q^1}_D} =0  \\
\{{\cal O}^{Q^{1}}_{p_{\mu_x}}(z') , {\cal O}^{Q^{1}}_{C_{e_1}}(z^{\prime\prime})\}
&=&{\cal O}^{Q^{1}}_{ \{ p_{\mu_x}(z'),C_{e_1}(z^{\prime\prime}) \}^{Q^1}_D} =0.
\end{eqnarray*}
Using that $ \{{\cal O}^{Q^{1}}_{p_{\mu_x}}(z'), {\cal O}^{Q^{1}}_{E^x}(y) \}_D^{Q^2} =0$ we can conclude $ \{ {\cal O}^{Q^{1\dots 2}}_{p_{\mu_x}}(z') ,{\cal O}^{Q^{1\dots 2}}_{E^x}(y) \}=0$ and likewise we obtain $\{{\cal O}^{Q^{1\dots 2}}_{p_{\mu_\varphi}}(z), {\cal O}^{Q^{1\dots 2}}_{p_{\mu_x}}(z')\}=0$. This yields
\begin{equation*}
  \{{\cal O}^{Q^{1\dots 2}}_{p_{\mu_\varphi}}(z), {\cal O}^{Q^{1\dots 2}}_{E^x}(y) \}_D^{Q^3}    
 =
 \{{\cal O}^{Q^{1\dots 2}}_{p_{\mu_\varphi}}(z), {\cal O}^{Q^{1\dots 2}}_{E^x}(y) \}.   
\end{equation*}
Now with similar iterative steps we can show that 
\begin{eqnarray*}
  \{{\cal O}^{Q^{1\dots 2}}_{p_{\mu_\varphi}}(z), {\cal O}^{Q^{1\dots 2}}_{E^x}(y) \}_D^{Q^3}    
& =&
 \{{\cal O}^{Q^{1\dots 2}}_{p_{\mu_\varphi}}(z), {\cal O}^{Q^{1\dots 2}}_{E^x}(y) \}   \\
 &=&{\cal O}^{Q^1}_{{\cal O}^{Q^2}_{\{p_{\mu_\varphi}(z),E^x(y)\}}}=0.
\end{eqnarray*}
These results yield 
\begin{eqnarray*}
 \{ {\cal O}^{Q^{1\dots 3}}_{p_{\mu_\varphi}}(z) ,{\cal O}^{Q^{1\dots 3}}_{E^x}(y) \}
 &=&
 {\cal O}^{Q^3}_{\{{\cal O}^{Q^{1\dots 2}}_{p_{\mu_\varphi}}(z), {\cal O}^{Q^{1\dots 2}}_{E^x}(y) \}_D^{Q^3}}=0. 
 \end{eqnarray*}
Then for the second Poisson bracket involving ${\cal O}^{Q^{1\dots 3}}_{p_{\mu_\varphi}}(z)$ we can perform exactly the same steps to end up with
\begin{eqnarray*}
 \{{\cal O}^{Q^{1\dots 3}}_{K_x}(x) , {\cal O}^{Q^{1\dots 3}}_{p_{\mu_\varphi}}(z)\}
& =&
{\cal O}^{Q^3}_{\{{\cal O}^{Q^{1\dots 2}}_{K_x}(x) , {\cal O}^{Q^{1\dots 2}}_{p_{\mu_\varphi}}(z)\}^{Q^3}_D} 
={\cal O}^{Q^3}_{\{{\cal O}^{Q^{1\dots 2}}_{K_x}(x) , {\cal O}^{Q^{1\dots 2}}_{p_{\mu_\varphi}}(z)\}} 
=
{\cal O}^{Q^{1\dots 3}}_{\{K_x(x),p_{\mu_\varphi}(z)\}}=0.
\end{eqnarray*}
This then leads to
\begin{equation*}
 \{{\cal O}^{Q^{1\dots 3}}_{K_x}(x), {\cal O}^{Q^{1\dots 3}}_{E^x}(y) \}_D^{Q^4}=
\{{\cal O}^{Q^{1\dots 3}}_{K_x}(x), {\cal O}^{Q^{1\dots 3}}_{E^x}(y) \}.   
\end{equation*}
The final step is to show that 
\begin{eqnarray}
\label{eq:PBdelta}
 \{{\cal O}^{Q^{1\dots 3}}_{K_x}(x), {\cal O}^{Q^{1\dots 3}}_{E^x}(y) \} =G\delta(x,y)   
\end{eqnarray}
This can be done with similar iterative steps. First we have
\begin{eqnarray*}
  \{{\cal O}^{Q^{1\dots 3}}_{K_x}(x), {\cal O}^{Q^{1\dots 3}}_{E^x}(y) \} 
  &=& {\cal O}^{Q^3}_{\{{\cal O}^{Q^{1\dots 2}}_{K_x}(x), {\cal O}^{Q^{1\dots 2}}_{E^x}(y) \}^{Q^3}_D}
  = {\cal O}^{Q^3}_{\{{\cal O}^{Q^{1\dots 2}}_{K_x}(x), {\cal O}^{Q^{1\dots 2}}_{E^x}(y) \}},
\end{eqnarray*}
where the reduction to the Poisson bracket in the last step can be done because as in the case of the Dirac brackets above there will be always Poisson brackets involved in the additional terms in the Dirac bracket that include the momenta $P_3$ which can be shown to vanish identically as above. Next we have 
\begin{eqnarray*}
  \{{\cal O}^{Q^{1\dots 2}}_{K_x}(x), {\cal O}^{Q^{1\dots 2}}_{E^x}(y) \} 
  &=& {\cal O}^{Q^2}_{\{{\cal O}^{Q^{1}}_{K_x}(x), {\cal O}^{Q^{1}}_{E^x}(y) \}^{Q^2}_D}
  = {\cal O}^{Q^2}_{\{{\cal O}^{Q^{1}}_{K_x}(x), {\cal O}^{Q^{1}}_{E^x}(y) \}}
\end{eqnarray*}
and using that 
\begin{eqnarray*}
  \{{\cal O}^{Q^{1}}_{K_x}(x), {\cal O}^{Q^{1}}_{E^x}(y) \} 
  &=& {\cal O}^{Q^1}_{ \{ K_x(x), E^x(y) \}^{Q^1}_D}
  = {\cal O}^{Q^1}_{\{ K_x(x), E^x(y) \}}=G\delta(x,y)
\end{eqnarray*}
we obtain directly 
\begin{eqnarray*}
 \{Q^5(x), P_5(y)\} &=&    \{{\cal O}^{Q^{1\dots 4}}_{K_x}(x), {\cal O}^{Q^{1\dots 4}}_{E^x}(y) \}
 ={\cal O}^{Q^4}_{\{{\cal O}^{Q^{1\dots 3}}_{K_x}(x), {\cal O}^{Q^{1\dots 3}}_{E^x}(y) \}_D^{Q^4}}=G\delta(x,y).
\end{eqnarray*}
~\\

Now it remains to show that the set of variables $(Q^I,P_I)$ for $I=5\dots N+6$ mutually commute with all $Q^I,P_I$ for $I=1\dots 4$. Here in general we will have Poisson brackets of quantities to which a different number of observable maps have been applied. We will demonstrate how the computation works in these cases for one specific example because the remaining ones can be computed similarly.
We start with 
\begin{eqnarray*}
\label{eq:PBQ5P3}
 \{Q^5(x),P_3(y)\}
 &=& \{ {\cal O}^{Q^{1\dots 4}}_{K_x}(x) , {\cal O}^{Q^{1\dots 2}}_{p_{\mu_x}}(y)\} \\
 &=&
 \sum\limits_{n=0}^\infty \int dz_1\dots  \int dz_n \{ \frac{(P_4(z_1)\dots P_4(z_n))}{n!} \nonumber \\
 && \{\dots \{\{ {\cal O}^{Q^{1\dots 3}}_{K_x}(x), Q_4(z_1)\},Q_4(z_2)\cdots,\}\},Q_4(z_n) \} , {\cal O}^{Q^{1\dots 2}}_{p_{\mu_x}}(y)\} \nonumber \\
 &=&
 \sum\limits_{n=0}^\infty \int dz_1\dots  \int dz_n  \frac{(P_4(z_1)\dots P_4(z_n))}{n!} \nonumber \\
&& \{\{\dots \{\{ {\cal O}^{Q^{1\dots 3}}_{K_x}(x), {\cal O}^{Q^{1\dots 3}}_{C_{\mu_{\varphi}}}(z_1)\},{\cal O}^{Q^{1\dots 3}}_{C_{\mu_\varphi}}(z_2)\cdots,\}\},{\cal O}^{Q^{1\dots 3}}_{C_{\mu_\varphi}}(z_n) \} , {\cal O}^{Q^{1\dots 2}}_{p_{\mu_x}}(y)\} \nonumber .
\end{eqnarray*}
Next we consider the nested Poisson bracket more in detail. We have
\begin{eqnarray}
\label{eq:nestedPB}
&& \{\{\dots \{\{ {\cal O}^{Q^{1\dots 3}}_{K_x}(x), {\cal O}^{Q^{1\dots 3}}_{C_{\mu_\varphi}}(z_1)\},{\cal O}^{Q^{1\dots 3}}_{C_{\mu_\varphi}}(z_2)\cdots,\}\},{\cal O}^{Q^{1\dots 3}}_{C_{\mu_\varphi}}(z_n) \} \nonumber \\
 &=& {\cal O}^{Q^{3}}_{ \{\{\dots \{\{ {\cal O}^{Q^{1\dots 2}}_{K_x}(x), {\cal O}^{Q^{1\dots 2}}_{C_{\mu_\varphi}}(z_1)\},{\cal O}^{Q^{1\dots 2}}_{C_{\mu_\varphi}}(z_2)\cdots,\}\},{\cal O}^{Q^{1\dots 2}}_{C_{\mu_\varphi}}(z_n) \}^{Q^3}_D} \nonumber \\
 &=& {\cal O}^{Q^{3}}_{ \{\{\dots \{\{ {\cal O}^{Q^{1\dots 2}}_{K_x}(x), {\cal O}^{Q^{1\dots 2}}_{f_{\mu_\varphi}}(z_1)\},{\cal O}^{Q^{1\dots 2}}_{f_{\mu_\varphi}}(z_2)\cdots,\}\},{\cal O}^{Q^{1\dots 2}}_{f_{\mu_\varphi}}(z_n) \}^{Q^3}_D} \nonumber \\
& =& {\cal O}^{Q^{3}}_{ \{\{\dots \{\{ {\cal O}^{Q^{1\dots 2}}_{K_x}(x), {\cal O}^{Q^{1\dots 2}}_{f_{\mu_\varphi}}(z_1)\},{\cal O}^{Q^{1\dots 2}}_{f_{\mu_\varphi}}(z_2)\cdots,\}\},{\cal O}^{Q^{1\dots 2}}_{f_{\mu_\varphi}}(z_n) \}},
\end{eqnarray}
where we used the gravitational degrees of freedom commute with all momenta whose corresponding observables are used as clocks in the individual observable maps to reduce the Dirac bracket to the corresponding Poison bracket in the last step. Now further we obtain
\begin{eqnarray*}
&& \{\{\dots \{\{ {\cal O}^{Q^{1\dots 2}}_{K_x}(x), {\cal O}^{Q^{1\dots 2}}_{C_{\mu_\varphi}}(z_1)\},{\cal O}^{Q^{1\dots 2}}_{C_{\mu_\varphi}}(z_2)\cdots,\}\},{\cal O}^{Q^{1\dots 2}}_{C_{\mu_\varphi}}(z_n) \} \\
 &=&{\cal O}^{Q^{1\dots 2}}_{\{\{\dots \{\{K_x(x),f_{\mu_\varphi}(z_1)\},f_{\mu_\varphi}(z_2)\},\dots,\}\},f_{\mu_\varphi}(z_n)\}}.
\end{eqnarray*}
We realize that for all values of $n\geq 0$ the nested Poisson bracket involves gravitational degrees of freedom only. For a more compact notation let us introduce the following abbreviation
\begin{eqnarray*}
 F^{\rm grav}(z_1,\dots,z_n):=\{\{\dots \{\{K_x(x),f_{\mu_\varphi}(z_1)\},f_{\mu_\varphi}(z_2)\},\dots,\}\},f_{\mu_\varphi}(z_n)\}, 
\end{eqnarray*}
the label 'grav' should indicate that the function depends on the gravitational degrees of freedom only. Note that the following steps will also hold if we generalize to the case where this function depends on gravitational as well as matter degrees of freedom.
Now we reinsert this function into \eqref{eq:nestedPB} and obtain
\begin{eqnarray*}
 && \{\{\dots \{\{ {\cal O}^{Q^{1\dots 3}}_{K_x}(x), {\cal O}^{Q^{1\dots 3}}_{C_{\mu_\varphi}}(z_1)\},{\cal O}^{Q^{1\dots 3}}_{C_{\mu_\varphi}}(z_2)\cdots,\}\},{\cal O}^{Q^{1\dots 3}}_{C_{\mu_\varphi}}(z_n) \} \\
 &=&{\cal O}^{Q^{3}}_{{\cal O}^{Q^1Q^2}_{ F^{\rm grav}(z_1,\dots,z_n)}} \\
 &=&\sum\limits_{m=0}^\infty \frac{1}{m!}\int ds_1\dots \int ds_m
 (P_3(s_1)\dots P_3(s_m))\{\{\dots\{{\cal O}^{Q^1Q^2}_{ F^{\rm grav}(z_1,\dots,z_n)}, Q_3(s_1)\}\dots\},Q_3(s_m)\} \\
 &=&\sum\limits_{m=0}^\infty \frac{1}{m!}\int ds_1\dots \int ds_m
 (P_3(s_1)\dots P_3(s_m))\{\{\dots\{{\cal O}^{Q^1Q^2}_{ F^{\rm grav}(z_1,\dots,z_n)}, {\cal O}^{Q^1Q^2}_{f_{\mu_x}}(s_1)\}\dots\},{\cal O}^{Q^1Q^2}_{f_{\mu_x}}(s_m)\} \\
 &=&\sum\limits_{m=0}^\infty \frac{1}{m!}\int ds_1\dots \int ds_m
 (P_3(s_1)\dots P_3(s_m)){\cal O}^{Q^1Q^2}_{\{\dots\{F^{\rm grav}(z_1,\dots,z_n), {f_{\mu_x}}(s_1)\}\dots\},{f_{\mu_x}}(s_m)\}}.
 \end{eqnarray*}
 If we reinsert this result into the \eqref{eq:PBQ5P3} and use that $\{P_3(x),P_3(y)\}=0$ finally we end up with 
 \begin{eqnarray*}
 \{Q^5(x),P_3(y)\} &=&
 \sum\limits_{n=0}^\infty \int dz_1\dots  \int dz_n  \frac{(P_4(z_1)\dots P_4(z_n))}{n!} \sum\limits_{m=0}^\infty \int ds_1\dots \int ds_m \frac{(P_3(s_1)\dots P_3(s_m))}{m!}\\
&& \{{\cal O}^{Q^1Q^2}_{\{\dots\{F^{\rm grav}(z_1,\dots,z_n), {\cal O}^{Q^1Q^2}_{f_{\mu_x}}(s_1)\}\dots\},{\cal O}^{Q^1Q^2}_{f_{\mu_x}}(s_m)\}}\, , \, {\cal O}^{Q^1Q^2}_{p_{\mu_x}}(y) \} \\
&=&\sum\limits_{n=0}^\infty \int dz_1\dots  \int dz_n  \frac{(P_4(z_1)\dots P_4(z_n))}{n!} \sum\limits_{m=0}^\infty \int ds_1\dots \int ds_m \frac{(P_3(s_1)\dots P_3(s_m))}{m!}\\
&& {\cal O}^{Q^1Q^2}_{\{\{\dots\{F^{\rm grav}(z_1,\dots,z_n), {\cal O}^{Q^1Q^2}_{f_{\mu_x}}(s_1)\}\dots\},{\cal O}^{Q^1Q^2}_{f_{\mu_x}}(s_m)\}\, ,\, {\cal O}^{Q^1Q^2}_{p_{\mu_x}}(y)\} } \\
&=& 0 ,
 \end{eqnarray*}
 where we used in last step that 
 \begin{equation*}
  \{\{\dots\{F^{\rm grav}(z_1,\dots,z_n), {f_{\mu_x}}(s_1)\}\dots\},{f_{\mu_x}}(s_m)\}\, ,\, {p_{\mu_x}}(y)\}  =0   
 \end{equation*}
 due to the fact that the involved iterated Poisson bracket depends on the gravitational degrees of freedom only and thus a further Poisson bracket with $p_{\mu_x}$ trivially vanishes and as a consequence also the application of the observables map ${\cal O
 }^{Q^{1\dots 2}}$. The remaining combinations of elementary phase space variables can be computed in a similar manner: some of them with the same and others with fewer iterations involved, which is why we considered the canonical pair $(Q^5,P_3)$ as a representative example.
\bibliographystyle{jhep}
\bibliography{refs}
\end{document}